\documentclass[letterpaper,12pt]{article}

\usepackage{color}

\usepackage{graphicx,cite}
\usepackage{amsmath}
\usepackage{amsfonts}
\usepackage{amssymb}

\begin{document}

\title{The $S_3$ flavour symmetry: Neutrino masses and mixings}

\author{\small{F. Gonz\'alez Canales$^{a),b)}$, A. Mondrag\'on$^{a)}$ and 
 M. Mondrag\'on$^{a)}$} \\
 \footnotesize{ $a)$~Instituto de F\'{\i}sica, Universidad Nacional Aut\'onoma de M\'exico,} \\
 \footnotesize{  Apdo. Postal 20-364, 01000, M\'exico D.F., M\'exico.} \\
 \footnotesize{ $b)$~Facultad de Ciencias de la Electr\'onica, Benem\'erita Universidad 
                 Aut\'onoma de Puebla,} \\
 \footnotesize{ Apdo. Postal 157, 72570, Puebla, Pue., M\'exico.} }

\maketitle 
 
\begin{abstract}
  In this work, we discuss the neutrino masses and mixings as the realization of an 
  $S_{3}$ flavour permutational symmetry in two models, namely the Standard Model and an extension 
  of the Standard Model with three Higgs doublets.  In the $S_3$ Standard Model, mass matrices of 
  the same generic form are obtained for the neutrinos and charged leptons when the $S_{3}$ 
  flavour symmetry is broken sequentially according to the chain $S_{3L} \otimes S_{3R} \supset 
  S_{3}^{diag} \supset S_{2}$. In the minimal $S_{3}$-symmetric extension of the Standard Model, 
  the $S_3$ symmetry is left unbroken, and the concept of flavour is extended to the Higgs sector 
  by introducing in the theory three Higgs fields which are $SU(2)$ doublets. In both models, the 
  mass matrices of the neutrino and charged leptons are reparametrized in terms of their 
  eigenvalues, and exact, explicit analytical expressions for the neutrino mixing angles as 
  functions of the masses of neutrinos and charged leptons are obtained. In the case of the $S_3$ 
  Standard Model, from a $\chi^{2}$ fit of the theoretical expressions of the lepton 
  mixing matrix to the values extracted from experiment, the numerical values of the neutrino 
  mixing angles are obtained in excellent agreement with experimental data.  In the $S_3$ 
  extension of the Standard Model, if two of the right handed neutrinos masses are degenerate, the 
  reactor and atmospheric mixing angles are determined by the masses of the charged leptons, 
  yielding $\theta_{23}$ in excellent agreement with experimental data, and $\theta_{13}$ 
  different from zero but very small.  If the masses of the three right handed neutrinos are 
  assumed to be different, then it is possible to get $\theta_{13}$ also in very good agreement 
  with experimental data.  We also show the branching ratios of some selected flavour changing 
  neutral currents (FCNC) process as well as the contribution of the exchange of a neutral flavour 
  changing scalar to the anomaly of the magnetic moment of the muon.
\end{abstract} 

 \section{Introduction}

  The observation of flavour oscillations of solar, atmospheric, reactor, and accelerator 
  neutrinos established that they have non-vanishing masses and mix among themselves, much like 
  the quarks do~\cite{Jung:2001dh,Mohapatra:2006gs,Altmann:2005ix,Smy:2003jf,Ahmad:2002ka,
  Ahmad:2002jz, Aharmim:2005gt,Fukuda:2002pe,Ashie:2005ik,Bemporad:2001qy,Araki:2004mb,
  Maltoni:2004ei,Schwetz:2005jr,GonzalezGarcia:2007ib,Apollonio:2002gd,K2005197,Elliott:2004hr,
  Seljak:2006bg,Elgaroy:2004rc,Lesgourgues:2006nd}. In these observations and experiments, the 
  differences of the squared neutrino masses as well as the neutrino mixing angles are measured. 
  These discoveries brought out very forcibly the need of extending the Standard Model (SM) in 
  order to accommodate in the theory the new data on neutrino physics in a consistent way, that 
  would allow for a unified and systematic treatment of the observed hierarchy of masses and 
  mixings of all fermions.  At the same time, the number of free parameters in the extended form 
  of the SM had to be drastically reduced in order to give predictive power to the theory.  These 
  two seemingly contradictory demands are met by a flavour symmetry under which the families 
  transform in a non-trivial fashion. The observed pattern of neutrino mixing and, in particular, 
  the non vanishing and sizable value of the reactor mixing angle strongly suggests a 
  flavour permutational symmetry $S_{3}$.
 
  The result of a combined analysis of all available neutrino oscillation data, including the 
  recent results from long-baseline $\nu_{\mu} \longrightarrow \nu_{e}$ searches at the Tokai to 
  Kamioka (T2K)~\cite{PhysRevLett.107.041801} and Double CHOOZ experiments~\cite{Abe:2011fz} as 
  well as the Main Injector Neutrino Oscillation Search (MINOS) 
  experiment~\cite{PhysRevLett.107.181802}, give the following values for the differences of the 
  squared neutrino masses and the mixing angles in the lepton mixing matrix,
  $U_{ _{PMNS} }$, at $1 \sigma$ confidence
  level~\cite{Schwetz:2011zk}:
   \begin{equation}\label{MCGCdatos}
    \begin{array}{l}
     \Delta m_{ 21 }^{ 2 } = 7.59^{ + 0.20 }_{ - 0.18 }  \times 10^{ -5 }~\textrm{eV}^{2}, \quad
     \begin{array}{l} 
      \Delta m_{ 31 }^{ 2 } = \left \{  
      \begin{array}{l} 
        -2.40_{-0.09}^{+0.08}  \times 10^{ -3 }~\textrm{eV}^{2}, \\ \\
        +2.50_{-0.16}^{+0.09} \times 10^{ -3 }~\textrm{eV}^{2}.
      \end{array} \right. 
     \end{array}
    \end{array}
   \end{equation}
   \begin{equation}
    \begin{array}{lll}
     \sin^{2} \theta_{12}^{ l } =  0.312_{-0.015}^{+0.017}, &
     \sin^{2} \theta_{23}^{ l } =  \left \{  
      \begin{array}{l} 
       0.52\pm 0.06 \\ \\
       0.52_{-0.07}^{+0.06} 
      \end{array} \right. , &
     \sin^{2} \theta_{13}^{ l } = \left \{  
      \begin{array}{l} 
       0.016_{-0.006}^{+0.008} \\ \\
       0.013_{-0.005}^{+0.007} 
      \end{array} \right. ,
    \end{array}
   \end{equation} 
   the upper (lower) row corresponds to inverted (normal) neutrino mass hierarchy, see also Gonzalez 
   Garcia {\it et  al}~\cite{GonzalezGarcia:2010er,springerlink:10.1134/S106377961104006X}
   and J. F. W. Valle {\it et al}~\cite{Schwetz:2011qt}. In fact, from the three angles needed to 
   describe the mixing of the neutrinos, the least known is $\theta_{13}^{ l } $. A global analysis of 
   the T2K~\cite{PhysRevLett.107.041801}, MINOS~\cite{PhysRevLett.107.181802} and 
   CHOOZ~\cite{Abe:2011fz} experiments yielded a non-vanishing value for the reactor mixing
   angle~\cite{Schwetz:2011qt,PhysRevD.84.053007}. Recently, the Daya Bay~\cite{An:2012eh} and RENO 
   experiments~\cite{Ahn:2012nd} found the following values for the reactor neutrino mixing angle:
   $\sin^{2} 2 \theta_{13}^{ l } = 0.092 \pm 0.016 \left( \textrm{stat} \right) \pm 0.005 
   \left( \textrm{syst} \right) $ which is equivalent to $ \theta_{13}^{ l } \simeq 8.8^{\circ} \pm
   0.8^{\circ} $ at $5.2 \; \sigma$ level, and $\sin^{2} 2 \theta_{13}^{ l } = 0.113 \pm 0.013 
   \left( \textrm{stat} \right) \pm 0.019 \left( \textrm{syst} \right) $ which is equivalent to 
   $\theta_{13}^{ l } \simeq 9.8^{\circ} $ at $4.9\; \sigma$ level~\cite{Ahn:2012nd}.
  
   In the last ten years, important theoretical advances have been made in the understanding of the 
   mechanisms for fermion mass generation and flavour mixing. The imposition of flavour symmetries
   in the Standard Model and its extensions strongly constrains the number of  free parameters in the 
   Yukawa couplings and gives rise to special forms of the fermions mass matrices with few free
   parameters and a number of texture 
   zeroes~\cite{Fritzsch:1999ee,Xing:2003zd,Canales:2011ug,Barranco:2010we}.
   For a recent review of flavour symmetry models
   see~\cite{Ishimori:2010au,RevModPhys.82.2701}.
   
   In the case of the Minimal $S_3$-Invariant Extension of the
   Standard
   Model~\cite{Kubo:2003iw,Kubo:2005sr,Felix:2006pn,Mondragon:2007af,Mondragon:2007nk,
     Mondragon:2007jx,Mondragon:2008gm}, the concept of flavour and
   generations is extended to the Higgs sector in such a way that all
   the matter fields -- Higgs, quarks, and lepton fields, including
   the right-handed neutrino fields-- have three species and transform
   under the flavour symmetry group as the three dimensional
   representation ${\bf 1} \oplus {\bf 2}$ of the permutational group
   $S_3$.  A model with more than one Higgs $SU(2)$ doublet has tree
   level flavour changing neutral currents whose exchange may give
   rise to lepton flavour violating processes and may also contribute
   to the anomalous magnetic moment of the muon.  An effective test of
   the phenomenological success of the model is obtained by verifying
   that all flavour changing neutral current processes and the
   magnetic anomaly of the muon, computed in the $S_3$--Invariant
   extended form of the Standard Model, agree with the experimental
   values.
   
   Another important application of the permutational group $S_{3}$ is
   the classification of physically equivalent mass matrices.
   Different mass matrices with texture zeroes located in different
   positions have exactly the same physical content if they are
   related by a similarity transformation in flavour
   space~\cite{Mondragon:1998gy,Barranco:2010we,Xing:2003zd,Fritzsch:1999ee}.
   If the invariants of the mass matrix are to be preserved, the
   elements on the diagonal can only exchange positions in the
   diagonal while the off diagonal elements can only exchange
   positions off the diagonal. So, the transformations matrices that
   define the similarity classes are the six elements of the three
   dimensional real representation of the group of permutations
   $S_{3}$~\cite{Branco:2007nn,Canales:2011ug}.
   
   The left-handed Majorana neutrinos naturally acquire their small
   masses through the type I seesaw mechanism 
   \begin{equation}\label{subibajadef}
     { \bf M }_{ \nu_{L} }  = 
     { \bf M }_{ \nu_{D} }  { \bf  M }_{ \nu_R }^{-1}   { \bf M }_{ \nu_{D} }^{T} ,
   \end{equation}
   where ${\bf M}_{ \nu_{D} }$ and $ { \bf M}_{ \nu_{R} }$ denote the Dirac and right handed Majorana 
   neutrino mass matrices, respectively.
   
   The mass matrices are diagonalized by bi-unitary transformations as
   \begin{equation}\label{unu}
    \begin{array}{l}
      { \bf U}_{iL}^{\dag} {\bf M}_{i} {\bf U}_{iR} 
      = \mbox{diag} (m_{i1}, m_{i2},m_{i3}), 
      \quad \textrm{and} \quad 
      { \bf U}_{\nu}^{T} {\bf M }_{ \nu_{L} } { \bf U}_{\nu} 
      = \mbox{diag} (m_{\nu_1},m_{\nu_2},m_{\nu_3}),
    \end{array}
   \end{equation}
   where $i= d, u,e $. The entries in the diagonal matrices may be
   complex, so the physical masses are their absolute values.

   The lepton flavour mixing matrices ${ \bf V}_{ _{PMNS} }$ arise
   from the mismatch between diagonalization of the mass matrices of
   charged leptons and left-handed neutrinos,
   \begin{equation}\label{ckm1}
    \begin{array}{ll}
     { \bf V}_{_{PMNS} } = { \bf U}_{eL}^{\dag} { \bf U}_{\nu } {\bf K}, 
    \end{array}
   \end{equation}
   where $ {\bf K}$ is the diagonal matrix of the Majorana phase
   factors. Therefore, in order to obtain the unitary matrices
   appearing in eq.~(\ref{ckm1}) and get predictions for the flavour
   mixing angles and CP violating phases, we should specify the mass
   matrices. Also, in the case of three neutrino mixing there are
   three CP violation rephasing invariants~\cite{Hochmuth:2007wq},
   associated with the three CP violating phases present in the ${ \bf
     V}_{ _{PMNS} }$ matrix. The rephasing invariant related to the
   Dirac phase, analogous to the Jarlskog invariant in the quark
   sector, is given by:
   \begin{equation}\label{InvJl}
    J_{ l } \equiv  \Im m \left[ V_{e1}^{ * } V_{ \mu 3 }^{ * } V_{ e3 } V_{ \mu 1 }\right] .
   \end{equation} 
   The rephasing invariant $J_{ l }$ controls the magnitude of CP
   violation effects in neutrino oscillations and is a directly
   observable quantity. The other two rephasing invariants associated
   with the two Majorana phases in the ${ \bf V}_{ _{PMNS} }$ matrix,
   can be chosen as: $ S_{1} \equiv \Im m \left[ V_{e1}V_{ e3 }^{ *
     }\right]$ and $S_{2} \equiv \Im m \left[ V_{e2}V_{ e3 }^{ *
     }\right]$. These rephasing invariants are not uniquely defined,
   but $J_{ l }$, $S_{1}$ and $S_{2} $ are relevant for the definition
   of the effective Majorana neutrino mass, $m_{ee}$, in the
   neutrinoless double beta decay.
   
   In the standard PDG parametrization~\cite{Nakamura:2010zzi}, the
   entries in the lepton mixing matrix are parametrized in terms of
   the mixing angles and phases. Thus, the mixing angles are related
   to the observable moduli of lepton $ {\bf V}_{ _{ PMNS } } $
   through the relations:
   \begin{equation}\label{angulosMezclas}
    \begin{array}{l}
     \sin^{2}{ \theta_{12}^{ l } } = \frac{ \left| V_{ e2 } \right|^{2} }{ 1 - \left| V_{ e3 } \right|^{2} }, \quad
     \sin^{2}{ \theta_{23}^{ l } } = \frac{ \left| V_{ \mu 3 } \right|^{2} }{  1 - \left| V_{ e3 } \right|^{2} }, 
     \quad  \sin^{2}{ \theta_{13}^{ l } } = \left| V_{ e3 }  \right|^{2}.
    \end{array}  
   \end{equation}
         
   The neutrino oscillations do not provide information about either
   the absolute mass scale or their nature, this is, if
   neutrinos are Dirac or Majorana particles~\cite{Camilleri:2008zz}.
   Thus, one of the most fundamental problems in neutrinos physics is
   the question of the nature of massive neutrinos. A direct way to
   reveal the nature of massive neutrinos is to investigate processes
   in which the total lepton number is not
   conserved~\cite{PhysRevD.76.116008}. The matrix elements for these
   processes are proportional to the effective Majorana neutrino
   masses, whose magnitudes squared are
    \begin{equation}\label{masa_eff.19}
     \begin{array}{l}
       \left| \langle m_{ll} \rangle \right|^{2} =  \sum_{j=1}^{3} m_{ \nu_{j} }^{ 2 } 
      \left| V_{ lj } \right|^{ 4 }   +  2 \sum_{j<k}^{3} m_{ \nu_{j} } m_{ \nu_{k} }
      \left| V_{ lj } \right|^{ 2 }  \left| V_{ lk } \right|^{ 2 }  
      \cos 2\left( w_{lj} - w_{lk} \right), 
      \; l = e, \mu, \tau,
    \end{array}
   \end{equation}
   where $m_{ \nu_{j} }$ are the neutrino Majorana masses, $V_{lj}$
   are the elements of the lepton mixing matrix, $ w_{lj} = \arg \left
     \{ V_{ lj } \right \}$; this term includes phases of both types,
   Dirac and Majorana.  The half-life of the neutrinoless double beta
   decay ($0\nu2 \beta$ decay) can be expressed as $\left[
     T_{1/2}^{0\nu2 \beta} \right]^{-1} = G_{_{0\nu}} \left| M_{0 \nu
     } \right|^{2} \left| \langle m_{ee} \rangle \right|^{2} $, where
   $G_{_{0\nu}}$ is a phase factor, $\left| M_{0 \nu } \right|$ is the
   isotope specific nuclear decay matrix element and $\left| \langle
     m_{ee} \rangle \right|$ is the magnitude of the effective
   Majorana neutrino masses defined in eq.(\ref{masa_eff.19}). Current
   experiment data yield only an upper bound for this quantity $\left|
     \langle m_{ee} \rangle \right| <
   0.3$~eV~\cite{KlapdorKleingrothaus:2000sn,KlapdorKleingrothaus:2004na}.

  \section{An $S_{3}$ flavour symmetry in the SM }
   
  In analogy with the work of A. Mondragon and E. Rodriguez Jauregui
  on the $S_{3}$ flavour symmetry in the quark sector of the Standard
  Model~\cite{Mondragon:1998gy,Mondragon:1999jt}, we will start by
  assuming the obvious, the one Higgs boson in the SM is an $SU(2)_{L}$
  doublet and, since it has no flavour, it can only be accommodated in
  a singlet representation of $S_{3}$. The mass term in the
  Lagrangian, obtained by taking the vacuum expectation value of the
  Higgs field in the quark and lepton Yukawa couplings, gives rise to
  mass matrices ${\bf M_d}$ , ${\bf M_u}$, ${\bf M}_l$ and ${\bf
    M_{\nu} }$~\cite{Barranco:2010we,Canales:2011ug};
   \begin{equation}\label{eq:21}
    {\cal L}_{Y} ={\bf \bar{q}}_{d,L}{\bf M}_{d}{\bf q}_{d,R} + {\bf\bar{q}}_{u,L}{\bf M}_{u}{\bf q}_{u,R}
     + {\bf \bar{L}}_{L}{\bf M}_{l}{\bf L}_{R} + {\bf \bar{\nu}}_{L}{\bf M}_{\nu} \left( {\bf \nu}_{L} \right)^{c}
     + h.c.
   \end{equation}
   If it is assumed that $S_{3}$ is an exact symmetry of the model,
   these mass matrices give mass only to the one fermion in each
   family that is assigned to the singlet representation of
   $S_{3}$. Hence, in a symmetry adapted basis, all entries in these
   matrices should vanish except for the one at the third row and
   third column. There is no mismatch between the diagonalization of
   the mass matrices of the charged leptons and neutrinos, or $d$-
   and $u$- type quarks, and, consequently there is no mixing of the
   flavour indices. Therefore, we propose, along with many other
   authors~\cite{Fritzsch:1977za,Mondragon:1998gy,Barranco:2010we,Fritzsch:1999ee,Kubo:2003iw},
   that the texture zeroes of the mass matrices of quarks and leptons 
   are the result of a flavour permutational symmetry $S_{3}$ and its
   spontaneous or explicit breaking.  In particular, the permutational
   $S_{3}$ flavour symmetry and its sequential explicit breaking
   allows us to justify using the same generic form for the mass
   matrices of all Dirac fermions~\cite{Barranco:2010we,
     Canales:2011ug}, this form is conventionally called a two texture
   zeroes form.  Some reasons to propose the validity of a matrix with
   two texture zeroes as a universal form for the mass matrices of all
   Dirac fermions in the theory are the following:
   \begin{enumerate}
   \item The idea of $S_{3}$ flavour symmetry and its explicit
     breaking has been succesfully realized as a mass matrix with two
     texture zeroes in the quark sector to describe the strong mass
     hierarchy of up and down type
     quarks~\cite{Mondragon:1998gy,Fritzsch:1977za, Pakvasa:1977in,
       Harari:1978yi, Fritzsch:1979zq}. Also, the numerical values of
     the mixing matrices of the quarks determined in this framework
     are in good agreement with the experimental
     data~\cite{Mondragon:1998gy,Barranco:2010we}.
   \item Since the mass spectrum of the charged leptons exhibits a
     hierarchy similar to the quark's one, it would be natural to
     consider the same $S_{3}$ symmetry and its explicit breaking to
     justify the use of the same generic form with two texture zeroes
     for the charged lepton mass matrix.
   \item As for the Dirac neutrinos, we have no direct information
     about the absolute values or the relative values of the Dirac
     neutrino masses, but the mass matrix with two texture zeroes can
     be obtained from an  $SO(10)$ Grand Unified Theory which describes
     well the data on masses and mixings of Majorana
     neutrinos~\cite{Buchmuller:2001dc,Bando:2003ei,Bando:2004hi}.
     Furthermore, from supersymmetry arguments, it would be sensible
     to assume that the Dirac neutrinos have a mass hierarchy similar
     to that of the u-quarks and it would be natural to take for the
     Dirac neutrino mass matrix also a matrix with two texture zeroes.
   \end{enumerate}

  \subsection{Mass matrices from the breaking of $S_{3L}\otimes S_{3R}$ }
  Some authors have pointed out that realistic Dirac fermion mass
  matrices result from the flavour permutational symmetry $S_{3L}
  \otimes S_{3R}$ and its spontaneous or explicit breaking according
  to the chain: $S_{3L} \times S_{3R} \supset
  S_{3}^{\textrm{diag}}\supset S_{2L} \times S_{2R} \supset
  S_{2}^{\textrm{diag}}$~\cite{Fritzsch:1999ee,Mondragon:1998gy,Mondragon:1999jt,Barranco:2010we,
  Canales:2011ug,Fritzsch:1977za,Kubo:2003iw,Xing:2010iu}. The group $S_{3}$ treats
  three objects symmetrically, while the structure ${\bf 1} \oplus
  {\bf 2}$ of its $3 \times 3$ matrix representations treats the
  generations differently and adapts itself readily to the
  hierarchical nature of the fermion mass spectra. As explained above,
  under exact $S_{3L} \otimes S_{3R}$ symmetry, the mass spectrum for
  either quark sector (up or down quarks) or leptonic sector (charged
  leptons or Dirac neutrinos) consists of one massive particle in a
  singlet irreducible representation and a pair of massless particles
  in a doublet irreducible representation of $S_{3L} \otimes S_{3R}$.
  In order to be more precise, in the case of exact $S_{3L}\otimes
  S_{3R}$ symmetry, and assuming that there is only one Higgs boson in
  the theory, this $SU(2)_{L}$ doublet can only be in a singlet
  representation of $S_{3}$, that is, it is a scalar with respect to
  the $S_{3}$ transformations. Hence, the corresponding mass matrices,
  ${\bf M}_{i3}$, are invariant with respect to a permutation of the
  family (columns) and flavour (rows ) indices, and take the form
   \begin{equation}\label{Mi3}
    {\bf M}_{i3}^{^{(W)}} = \frac{ \left( 1 - \Delta \right)  }{ 3 m_{i3} } 
    \left( \begin{array}{ccc}
     1 & 1 &  1 \\
     1 & 1 &  1 \\
     1 & 1 &  1 
    \end{array}\right)_{W}, \qquad i = u, d, l, \nu_{_D}
   \end{equation}
   the subindex $W$ stands for weak basis. In order to make explicit
   the assignment of particles to irreducible representations of
   $S_{3}$, it will be convenient to make a change of basis from the
   weak basis to a symmetry adapted or hiererchical basis by means of
   the unitary matrix that diagonalizes the matrix ${\bf M}_{i3}$,
   \begin{equation}
    {\bf M}_{i3}^{^{(H)}} = {\bf U}^{\dagger} {\bf M}_{i3}^{^{(W)}} {\bf U}
   \end{equation}
   where 
   \begin{equation}
    {\bf U} = \frac{1}{ \sqrt{6} } 
    \left( \begin{array}{ccc}
     \sqrt{3} & 1 & \sqrt{2} \\
     -\sqrt{3} & 1 & \sqrt{2} \\
     0 & -2 & \sqrt{2}
    \end{array} \right) \quad \textrm{and} \quad
     {\bf M}_{i3}^{^{(H)}} = m_{i3}  
    \left( \begin{array}{ccc}
     0 & 0 &  0 \\
     0 & 0 &  0 \\
     0 & 0 &  1 - \Delta_{i} 
    \end{array}\right)_{H}.
   \end{equation}
   In the Standard Model with the $S_{3}$ symmetry, masses for the
   first and second families are generated if we add the terms ${\bf
     M}_{i2}$ and ${\bf M}_{i1}$ to ${\bf M}_{i3}$. The term ${\bf
     M}_{i2}$ breaks the permutational symmetry $S_{3L} \otimes
   S_{3R}$ down to $S_{2L} \otimes S_{2R}$ and mixes the singlet and
   doublet representation of $S_{3}$, while the term ${\bf M}_{i1}$
   transform as the mixed symmetry term of the doublet complex
   tensorial representation of the $S_{3}^{\textrm{diag}}$ diagonal
   subgroup of $S_{3L} \otimes S_{3R}$. Thus, {
     \small \begin{equation}\label{Mi1} {\bf M}_{i2}^{^{(W)}} = \frac{
         m_{i3} }{3} \left( \begin{array}{ccc}
           \alpha_{i}  & \alpha_{i}  &  \beta_{i}  \\
           \alpha_{i}  & \alpha_{i}  &  \beta_{i}  \\
           \beta_{i} & \beta_{i} & - 2 \beta_{i}
    \end{array}\right)_{W} \, \textrm{and} \,
    {\bf M}_{i1}^{^{(W)}} = \frac{ m_{i3}  }{ \sqrt{3} } 
    \left( \begin{array}{ccc}
     A_{i1}  & i A_{i2}  &  - A_{i1} - i  A_{i2}  \\
     - i A_{i2}  & \alpha_{i}  &  \beta_{i}  \\
     \beta_{i}  & \beta_{i}  &  - 2 \beta_{i}  
    \end{array}\right)_{W}. 
   \end{equation} }
   Finally, adding the mass matrices eqs.~(\ref{Mi3})-(\ref{Mi1}), we get the mass matrix 
   ${ \bf M}_{i}$ in the 
   weak basis. Then, in a symmetry adapted basis ${ \bf M}_{i}$ takes the form 
   \begin{equation}\label{eq:2.11}
    { \bf M}_{i} = m_{i3} 
    \left( \begin{array}{ccc}
     0 & A_{i} & 0 \\
     A_{i}^{*} & B_{i} & C_{i} \\
     0 & C_{i} & D_{i} 
     \end{array} \right)_{H} ,\qquad i = u, d, l, \nu_{_D}
   \end{equation}
   where $A_{i} = |A_{i}|e^{i\phi_{i}}$, $B_{i} = -\triangle_{i} +
   \delta_{i}$ and $D_{i} = 1 - \delta_{i}$. From the strong hierarchy
   of the masses of the Dirac fermions, $m_{i3} >> m_{i2} > m_{i1}$,
   we expect $1-\delta_{i}$ to be very close to unity. The Hermitian
   mass matrix (\ref{eq:2.11}) may be written in terms of a real
   symmetric matrix ${\bf \overline{M} }_{ i }$ and a diagonal matrix of
   phases ${\bf P}_{ i }\equiv \textrm{diag}\left[ 1, e^{ i\phi_{ i }
     }, e^{ i\phi_{ i } } \right]$ as follows:
   \begin{equation}\label{Polar:FT}
    {\bf M}_{ i } = {\bf P}^{\dagger}_{ i }{\bf \bar{M} }_{ i } {\bf P}_{ i }, \qquad i = u, d, l, \nu_{_D}.
   \end{equation}
   
   Each possible symmetry breaking pattern is now characterized by the
   flavour symmetry breaking parameter ${Z_{i}}^{1/2}$, which is
   defined as the ratio {\small ${Z_{i}}^{1/2} = \frac{ \left( {M_{i}}
       \right)_{23} }{ \left({M_{i}} \right)_{22} }$}. This ratio
   measures the mixing of the singlet and doublet irreducible
   representations of $S_{3}$. The small parameter $\delta_{i}$ is a
   function of the flavour symmetry breaking parameter
   ${Z_{i}}^{1/2}$~\cite{Mondragon:1998gy,Barranco:2010we,Canales:2011ug}.
   
   Thus, we obtain a universal form for the mass matrices of all Dirac
   fermions in the theory. But in the Standard Model and its
   extensions considering a mass term for left-handed neutrinos purely
   of Dirac nature is not theoretically favored, because it cannot
   explain naturally why neutrinos are much lighter than the charged
   leptons. Thus, we assume that the neutrinos have Majorana masses and
   acquire their small masses through the type~I seesaw mechanism .

  \subsection{Classification  of mass matrices with texture zeroes in equivalence classes}
  In this section we make a classification of mass matrices with
  texture zeroes in terms of similarity clases \cite{Canales:2011ug}.  The similarity
  classes are defined as follows: Two matrices ${\bf M}$ and ${\bf
    M}'$ are similar if there exists an invertible matrix ${\bf T}$
  such that ${\bf M}' = {\bf T} {\bf M} {\bf T}^{-1}$ or ${\bf M}' =
  {\bf T}^{-1} {\bf M} {\bf T}$. The equivalence classes associated
  with a similarity transformation are called similarity classes.
  Another way to see the similarity classes is that the matrices that
  satisfy the similarity transformation have the same invariants: 
  trace, determinant and $\chi \equiv \frac{1}{2} \left( \textrm{Tr}
    \left \{ {\bf M}^{2} \right \} - \textrm{Tr} \left \{ {\bf M}
    \right \}^{2} \right)$. Therefore, all matrices in a class of
  similarity have the same eigenvalues​​, since all have the same
  characteristic polynomial, given by $ \lambda_{i}^{3} - \textrm{Tr}
  \left \{ {\bf M}^{2} \right \} \lambda_{i}^{2} - \chi \lambda_{i}
  -\textrm{det} \left \{ {\bf M} \right \} = 0$.
   
   Now, from the  most general form of the symmetric and  Hermitian mass matrices of $3\times3$:
   \begin{equation}\label{matrices:Herm:Sime}
    {\bf M}^{ ^{\textrm{s}} } =
    \left( \begin{array}{ccc}
     g & a & e \\
     a & b & c \\
     e & c & d  
    \end{array} \right)
    \quad \textrm{and} \quad
    {\bf M}^{ ^{\textrm{h} } } =
    \left( \begin{array}{ccc}
     g & a & e \\
     a^{*} & b & c \\
     e^{*} & c^{*} & d  
    \end{array} \right),
   \end{equation}
   we can see that only six of the nine elements of these matrices are
   independent of each other. Therefore, the similarity transformation
   is realized as the permutation of the six independent elements in
   the nine entries of the mass matrices. But if we want to preserve
   the invariants, the elements on the diagonal can only exchange
   positions on the diagonal, while the off-diagonal elements can only
   exchange positions outside the diagonal. Thus, all these operations
   reduce to the permutations of three objects.  So it is natural to
   propose as transformation matrices ${\bf T}$ in the similarity
   clases, the six elements of the real representation of the group of
   permutations $S_ {3}$ which are: { \small
     \begin{equation}\label{Base:completa}
    \begin{array}{l}
     {\bf T} \left( A_0 \right)=
      \left( \begin{array}{ccc}
       1 & 0 & 0 \\
       0 & 1 & 0 \\
       0 & 0 & 1
      \end{array}\right), \;
      {\bf T} \left(A_1\right)=
      \left( \begin{array}{ccc}
       0 & 1 & 0 \\
       1 & 0 & 0 \\
       0 & 0 & 1
      \end{array}\right), \;
      {\bf T} \left(A_2\right)=
      \left( \begin{array}{ccc}
       0 & 0 & 1 \\
       0 & 1 & 0 \\
       1 & 0 & 0
      \end{array}\right),\\\\
      {\bf T} \left(A_3\right)=
      \left(  \begin{array}{ccc}
       1 & 0 & 0 \\
       0 & 0 & 1 \\
       0 & 1 & 0
      \end{array}\right), \;
      {\bf T} \left(A_4\right)=
      \left(  \begin{array}{ccc}
       0 & 1 & 0 \\
       0 & 0 & 1 \\
       1 & 0 & 0
      \end{array}\right), \;
      {\bf T} \left(A_5\right)=
      \left(  \begin{array}{ccc}
       0 & 0 & 1 \\
       1 & 0 & 0 \\
       0 & 1 & 0
      \end{array}\right).
     \end{array} 
   \end{equation} }
   Then, we get the classification of mass matrices with texture zeroes, which is shown in the 
   table~1. In this table, the "${\star}$" and "$\times$" denote an arbitrary non-vanishing matrix 
   element on the diagonal and off-diagonal entries, respectively.  We recall the rule for counting the 
   texture zeroes 
   in a mass matrix:  two texture zeroes off-diagonal counts as one zero, while one on the diagonal 
   counts as one~\cite{Fritzsch:1999ee}. 
  \begin{table}
  \begin{center}
   \caption{ Matrix with two (left)  and one (right) texture zeroes.} \label{tabla3}
   \begin{tabular}{|c|c||c|} \hline
    {\footnotesize Class } &  {\footnotesize Textures } &   {\footnotesize Textures } \\ \hline
    {\footnotesize I} &  {\scriptsize $ \begin{array}{l} 
 \left( \begin{array}{ccc}
  0 & \times & 0 \\
  \times & \star & \times \\
  0 & \times & \star 
 \end{array}  \right)
 \left( \begin{array}{ccc}
   0 & 0 & \times \\
   0 & \star & \times \\
   \times & \times & \star  
 \end{array}  \right)
 \left( \begin{array}{ccc}
   \star & 0 & \times \\
    0 & 0 & \times \\
    \times & \times & \star   
 \end{array}  \right)
  \\
  \left( \begin{array}{ccc}
  \star & \times & \times \\
   \times & 0 & 0 \\
   \times & 0 & \star 
 \end{array}  \right)
 \left( \begin{array}{ccc}
  \star & \times & 0 \\
  \times & \star & \times \\
   0 & \times & 0 
 \end{array}  \right)
 \left( \begin{array}{ccc}
   \star & \times & \times \\
   \times & \star & 0 \\
   \times & 0 & 0 
 \end{array}  \right)
\end{array} $  }
  & 
{\scriptsize $\begin{array}{l} 
   \left( \begin{array}{ccc}
  0 & \times & \times \\
  \times & \star & \times \\
  \times & \times & \star 
 \end{array}  \right)
 \left( \begin{array}{ccc}
  \star & \times & \times \\
  \times & 0 & \times \\
  \times & \times & \star  
 \end{array}  \right) \\
 \left( \begin{array}{ccc}
   \star & \times & \times \\
   \times & \star & \times \\
   \times & \times & 0 
 \end{array}  \right)
  \end{array} $ } \\ \hline
  {\footnotesize II } & {\scriptsize $ \begin{array}{l} 
 \left( \begin{array}{ccc}
  0 & \times & \times \\
  \times & \star & 0 \\
  \times & 0 & \star 
 \end{array}  \right)
 \left( \begin{array}{ccc}
   \star & \times & 0 \\
   \times & 0 & \times \\
    0 & \times & \star 
 \end{array}  \right)
 \left( \begin{array}{ccc}
   \star & 0 & \times \\
    0 & \star & \times \\
    \times & \times & 0  
  \end{array} \right)
  \end{array} 
 $ }
  & 
{\scriptsize $\begin{array}{l} 
   \left( \begin{array}{ccc}
  \star & 0 & \times \\
   0 & \star & \times \\
  \times & \times & \star 
 \end{array}  \right)
 \left( \begin{array}{ccc}
  \star & \times & \times \\
  \times & \star & 0 \\
  \times &  0 & \star  
 \end{array}  \right) \\
 \left( \begin{array}{ccc}
   \star & \times & 0 \\
   \times & \star & \times \\
   0 & \times & \star 
 \end{array}  \right)
  \end{array} $ } \\ \hline
  {\footnotesize III } & {\scriptsize $ \begin{array}{l} 
 \left( \begin{array}{ccc}
  0 & \times & \times \\
  \times & 0 & \times \\
  \times & \times & \star 
 \end{array}  \right)
 \left( \begin{array}{ccc}
   0 & \times & \times \\
   \times & \star & \times \\
   \times & \times & 0 
 \end{array}  \right)
 \left( \begin{array}{ccc}
   \star & \times & \times \\
   \times & 0 & \times \\
   \times & \times & 0  
  \end{array} \right)
  \end{array} 
 $ }
  & 
  \\ \hline
 {\footnotesize IV } & {\scriptsize $ \begin{array}{l} 
 \left( \begin{array}{ccc}
  \star & 0 & 0 \\
   0 & \star & \times \\
   0 & \times & \star 
 \end{array}  \right)
 \left( \begin{array}{ccc}
   \star & 0 & \times \\
    0 & \star & 0 \\
   \times & 0 & \star 
 \end{array}  \right)
 \left( \begin{array}{ccc}
   \star & \times & 0 \\
   \times & \star & 0 \\
    0 & 0 & \star   
  \end{array} \right)
  \end{array} 
 $ } &  \\ \hline
\end{tabular}
\end{center}
\end{table} 

  \subsection{The mass matrix for left-handed neutrinos}\label{seccion2.4}
  The left-handed Majorana neutrinos acquire their small masses
  through the type~I seesaw mechanism , eq.~(\ref{subibajadef}). The
  form of ${\bf M}_{ \nu_{D} }$ is given in eq.~(\ref{eq:2.11}), which
  is a Hermitian matrix with two texture zeroes of class~I. From the
  conjecture of a universal $S_{3}$ flavour symmetry in a unified
  treatment of all fermions, it is natural to take for ${\bf M}_{
    \nu_R }$ also a matrix with two texture zeroes of class I, non
  Hermitian but symmetric. Let us further assume that the phases in
  the entries of ${\bf M}_{ \nu_{ R} }$ may be factorized out
  as~\cite{Barranco:2010we,Canales:2011ug}: ${\bf M}_{ \nu_{ _R } } =
  { \bf R } { \bf \bar{M} }_{ \nu_{ _R } } { \bf R }$, where ${ \bf R
  } \equiv \textrm{diag}\left[ e^{ - i\phi_{ c }}, e^{ i\phi_{ c } },
    1 \right]$ with $\phi_{ c } \equiv \arg \left \{ c_{ \nu_{ _R } }
  \right \}$ and
   \begin{equation}
    { \bf \bar{M} }_{ \nu_{ _R } } = \left( \begin{array}{ccc} 
     0 &  a_{ \nu_{ _R } }  & 0  \\
     a_{ \nu_{ _R } }   & | b_{ \nu_{ _R } } |    &| c_{ \nu_{ _R } } | \\
     0 &  | c_{ \nu_{ _R } }  | & d_{ \nu_{ _R } }   
    \end{array}\right).
   \end{equation}
   Then, the mass matrix of the left-handed Majorana neutrinos has
   also the same generic form with two texture zeroes of class I:
   \begin{equation}\label{seesaw:F}
     { \bf M }_{ \nu_{ _L } } =  
     \left( \begin{array}{ccc} 
      0 &  a_{ \nu_{ _L } } & 0  \\
      a_{ \nu_{ _L } } & b_{ \nu_{ _L } } & c_{ \nu_{ _L } }  \\
      0 & c_{ \nu_{ _L } } & d_{ \nu_{ _L } }   
     \end{array}\right),
   \end{equation}
   where 
   {\small \begin{equation}
    \begin{array}{l}
     a_{ \nu_{ _L } } = \frac{ | a_{ \nu_{ _D } } |^{2} }{  a_{ \nu_{ _R } } }, \,
     d_{ \nu_{ _L } } = \frac{ d_{ \nu_{ _D } }^{2} }{ d_{ \nu_{ _R } } }, \,
      c_{ \nu_{ _L } } = \frac{ c_{ \nu_{ _D } } d_{ \nu_{ _D } } }{ d _{ \nu_{ _R } } }+  
      \frac{ | a_{ \nu_{ _D } } | }{   | a_{ \nu_{ _R } } |}
      \left ( c_{ \nu_{ _D } }e^{- i \phi_{ \nu_{ D } } } - \frac{ |c_{ \nu_{ _R } }| d_{ \nu_{ D } }  }{ 
      d_{ \nu_{ _R } } } e^{ i \left( \phi_{ c } - \phi_{ \nu_{ D } }
      \right) } \right) \, \\
      b_{ \nu_{ _L } } = \frac{ c_{ \nu_{ _D } }^{ 2 } }{ d_{ \nu_{ _R } } }  + 
      \frac{ | c_{ \nu_{ _R } }|^{2} -  | b_{ \nu_{ _R } }|  d_{ \nu_{ _R } } }{  d_{ \nu_{ _R } } } 
      \frac{ | a_{ \nu_{ _D } } |^{ 2 } }{  a_{ \nu_{ _R } }^{2} } e^{ i 2\left( \phi_{ c } -
      \phi_{ \nu_{ D } } \right)} + 2 \frac{ | a_{ \nu_{ _D } }| }{ |a_{ \nu_{ _R } } | } \left( b_{ \nu_{ _D } } 
      e^{- i \phi_{ \nu_{ D } } } - \frac{c_{ \nu_{ _D } } |c_{ \nu_{ _R } }| }{ d_{ \nu_{ _R } }  }  
      e^{ i \left( \phi_{ c } - \phi_{ \nu_{ D } }   \right) } \right)
    \end{array}
  \end{equation} }  
The elements $a_{ \nu_{ _L } }$ and $d_{ \nu_{ _L } }$ are real, while $b_{ \nu_{ _L } }$ and $c_{ \nu_{ _L } }$ 
are complex. Therefore, the form of mass matrices with two texture zeroes is invariant under the action of 
the seesaw mechanism of type~I~\cite{Nishiura:1999yt,Fritzsch:1999ee,Xing:2003zd,Barranco:2010we}. 

It may also be noticed that, if we set $b_{ \nu_{ _R } } = 0$ or/and
$c_{ \nu_{ _R } } = 0$, the resulting expression for ${ \bf M}_{ \nu_{
    _L } }$ still has two texture zeroes. Therefore, ${ \bf M}_{ \nu_{
    _L } }$ also has two texture zeroes when ${ \bf M}_{ \nu_{ _R }
}$ has two, three or four texture zeroes (the last two cases are
called Fritzsch textures)~\cite{Barranco:2010we,Canales:2011ug}. In
fact, the information on the total number of texture zeroes present in
the mass matrix $ { \bf M}_{ \nu_{ _R } }$, can be obtained from the
elements $\left( 2,2 \right)$ and $\left( 2,3 \right)_{23}$ of the
left-handed neutrinos mass matrix.

From the previous analysis, the matrix ${ \bf M}_{ \nu_{ _L } }$ has
two non-ignorable phases which are $ \phi_{1} \equiv \arg \left\{ b_{
    \nu_{ _L } } \right \} $ and $\phi_{2} \equiv \arg \left\{ c_{
    \nu_{ _L } } \right \}$. However, to describe the phenomenology of
neutrinos masses and mixing, only one phase in $ { \bf M}_{ \nu_{ _L }
}$ is required. Therefore, without loss of generality, we may choose
$\phi_{1}=2\phi_{2}= 2\varphi$~\cite{Barranco:2010we,Canales:2011ug}.
In this case the analysis simplifies, since the phases in ${ \bf M}_{
  \nu_{ _L } }$ may be factorized out as
   \begin{equation}\label{Polar:FT2}
    { \bf M}_{ \nu_{ _L } } = { \bf Q} { \bf \bar{M} }_{ \nu_{ _L } } { \bf Q},
   \end{equation} 
   where ${ \bf Q}$ is a diagonal matrix of phases {\small ${ \bf Q}
     \equiv \textrm{diag}\left[ e^{ -i \varphi }, e^{ i \varphi }, 1
     \right] $ } and ${ \bf \bar{M} }_{ \nu_{ _L } }$ is a real
   symetric matrix. Then, the matrix ${ \bf M}_{ \nu_{ _L } }$, can be
   diagonalized by a unitary matrix of the form ${ \bf U}_{ \nu }
   \equiv { \bf Q} { \bf O}_{\nu}$, where $ {\bf O_{\nu} }$ is an real
   orthogonal matrix that diagonalizes the real symetric matrix ${\bf
     \bar{M} }_{ \nu_{ _L } }$.

  \subsection{Mass matrix as function of the fermion masses }
   The real symetric matrix ${ \bf \bar{M} }_{ i }$, eqs.~(\ref{Polar:FT}) and~(\ref{Polar:FT2}), may be brought to 
   diagonal form by the transformation,
   ${ \bf \bar{M} }_{ i } = {\bf O }_{ i } \textrm{diag} \left \{ m_{i1} , m_{i2} , m_{i3}  \right \}
    {\bf O }^{ T }_{ i } $, where the $m_{ i }$'s are the eigenvalues of $ {\bf M}_{ i }$ and  
    ${\bf O }_{ i }$ is a real orthogonal matrix, with $i = u, \; d, \; l, \; \nu_{ _L }$. 
    Now, computing the invariants of the real symetric matrix ${ \bf \bar{M} }_{ i }$, we may express the real 
    parameters $a_{ i }$, $b_{ i }$, $c_{ i }$ and $d_{ i }$ occuring in  eqs.~(\ref{Polar:FT}) and~(\ref{Polar:FT2})
    in terms of the mass eigenvalues~\cite{Barranco:2010we,Canales:2011ug} as:
   \begin{equation}\label{S3Ceros:63}
    a_{ i }^{2} = - \frac{ m_{i1} m_{i2} m_{i3} }{ d_{ i } }, \;
    b_{ i } = m_{i1} +m_{i2} + m_{i3} - d_{ i }, \;
    c_{ i }^{2} =  \frac{ \left( d_{ i } - m_{i1} \right ) \left( d_{ i } - m_{i2}\right)
    \left( m_{i3} - d_{ i } \right ) }{ d _{ i } } .
   \end{equation}
   From the condition that $a_{ i }$, $b_{ i }$, $c_{ i }$ and $d_{ i }$ are real, we determine the allowed region of
   $d_{ i }$~\cite{Matsuda:2006xa}. In other words, all elements of the matrix ${ \bf \bar{M} }_{ i }$ must be real and 
   depending on which eigenvalue is chosen as negative, the parameter $d_{ i }$ must satisfy one of the following 
   conditions:
   \begin{center}
    \begin{tabular}{l|l}
     Normal hierarchy  &  Inverted hierarchy        \\
     $m_{i3} > d_{i} > m_{i2}, \quad \textrm{for} \quad  m_{i1} = -|m_{i1}|$, &
     $m_{i2} > d_{i} > m_{i3}, \quad \textrm{for} \quad  m_{i1} = -|m_{i1}|$, \\
	 $m_{i3} > d_{i} > m_{i1}, \quad \textrm{for} \quad  m_{i2} = -|m_{i2}|$, &
	 $m_{i1} > d_{i} > m_{i3}, \quad \textrm{for} \quad  m_{i2} = -|m_{i2}|$, \\
     $m_{i2} > d_{i} > m_{i1}, \quad \textrm{for} \quad  m_{i3} = -|m_{i3}|$, &
     $m_{i2} > d_{i} > m_{i1}, \quad \textrm{for} \quad  m_{i3} = -|m_{i3}|$.    	          
    \end{tabular}
   \end{center}
   In this  way, for a  normal hierarchy and taking $m_{i2} = -|m_{i2}|$, we get the ${ \bf \bar{M} }_{ i }$ 
   matrix~$(i=u,d,l,\nu_{ _L })$,  reparametrized in terms of its eigenvalues and the parameter $\delta_{ i } $ as
   \begin{equation}\label{T_fritzsch:ev}
    { \bf \bar{M} }_{ i } =  \left( \begin{array}{ccc} 
     0 & \sqrt{ \frac{ \widetilde{ m }_{ i1 }   
     \widetilde{ m }_{ i2 } }{ 1 -  \delta_{ i } } } & 0 \\ 
     \sqrt{ \frac{ \widetilde{ m }_{ i1 }   \widetilde{ m }_{ i2 } }{  1 - \delta_{ i } } } & 
     \widetilde{ m }_{ i1 }  - \widetilde{ m }_{ i2 } +  \delta_{ i } & 
     \sqrt{ \frac{\delta_{ i } }{ ( 1 - \delta_{ i } ) } f_{ i1 }  f_{ i2 } }   \\ 
     0 & \sqrt{ \frac{ \delta_{ i } }{ ( 1 - \delta_{ i } ) } f_{ i1 } f_{ i2 } } &  1 - \delta_{ i }
    \end{array}\right),
   \end{equation}
   where  $\widetilde{m}_{i1} = \frac{ m_{i1} }{ m_{i3} }$, $\widetilde{m}_{i2} = \frac{ | m_{i2} | }{ m_{i3} }$,
   $ f_{i1}=1-\widetilde{m}_{i1}-\delta_{i}$ and $f_{i2} =1+ \widetilde{m}_{i2}  - \delta_{i}$.

   The small parameters $\delta_{i}$, which are also functions of the
   mass ratios and the flavour symmetry breaking parameters
   $Z^{1/2}_{i}$, are obtained as the solution of the cubic equation $(
   1 - \delta_{ i } ) (\widetilde{ m }_{ i1 } - \widetilde{ m }_{ i2 }
   + \delta_{ i } )^{2} Z_{ i } = \delta_{ i } f_{ i1 } f_{ i2 }$. The
   solution of the cubic equation that vanishes when $Z_{i}$ vanishes
   may be written as
   \begin{equation}
    \delta_{i} = \frac{ Z_{ i } }{ Z_{ i } + 1 } 
    \frac{ \left( \widetilde{m}_{i2} - \widetilde{m}_{i1} \right)^{2} }{ W _{i}\left( Z_{ i } \right) },
   \end{equation}
   where $W _{i}\left( Z \right)$ is the product of the two roots of the  cubic equation that do not vanish 
   when $Z_{i}$ vanishes
   {\small \begin{equation}
    \begin{array}{l}
     W _{i}\left( Z \right)  = \left[ p^{3}_{i} + 2 q^{2}_{i} + 2q \sqrt{ p^{3}_{i} +
     q^{2}_{i} } \right]^{ \frac{ 1 }{ 3 } } - | p_{i} |   
     + \left[ p^{3}_{i} + 2 q^{2}_{i} - 2q_{i} \sqrt{ p^{3}_{i} + q^{2}_{i} } \right]^{ \frac{ 1 }{ 3 } } + \\
     - \frac{1}{3} \left( \left[ q_{i} + \sqrt{ p^{3}_{i} + q^{2}_{i} } \right]^{ \frac{ 1 }{ 3 } } 
     + \left[ q_{i}- \sqrt{ p^{3}_{i} + q^{2}_{i} } \right]^{ \frac{ 1 }{ 3 } } \right) 
       \left( Z_{i} \left( 2 \left( \widetilde{m}_{i2} - \widetilde{m}_{i1}\right) + 1 \right)  
       \right. \\ \left.
      + \left( \widetilde{m}_{i2} - \widetilde{m}_{i1}\right) + 2 \right) 
     + \frac{1}{9} \left( Z_{i} \left( 2 \left( \widetilde{m}_{i2} - \widetilde{m}_{i1}\right) + 1 \right) 
     + \left( \widetilde{m}_{i2} -  \widetilde{m}_{i1}\right) + 2 \right)^{2},
    \end{array}
   \end{equation} }
   with 
   \[
    \begin{array}{l}
    p_{i} = -\frac{1}{3} \frac{ Z_{ i } \left( Z_{i} 
     \left( 2 \left( \widetilde{m}_{i2} - \widetilde{m}_{i1}\right) + 1 \right) 
     + \widetilde{m}_{i2} - \widetilde{m}_{i1} + 2 \right)^{2} }{ Z_{ i } + 1 } 
     + \frac{ \left[ Z_{i} \left( \widetilde{m}_{i2} - \widetilde{m}_{i1} \right) 
     \left( \widetilde{m}_{i2} - \widetilde{m}_{i1} + 2 \right) 
     \left(1 + \widetilde{m}_{i2} \right) \left( 1 - \widetilde{m}_{i1} \right) \right] }{ Z_{ i } + 1 }
    \end{array}
   \]
   and 
   \[
    \begin{array}{l}
     q_{i} = \frac{1}{6} \frac{ \left[ Z_{i} \left( \widetilde{m}_{i2} 
     - \widetilde{m}_{i1}  \right) \left( \widetilde{m}_{i2} - \widetilde{m}_{i1} + 2 \right) 
     \left(1 + \widetilde{m}_{i2} \right) \left( 1 - \widetilde{m}_{i1} \right) \right]
     \left( Z_{i} \left( 2 \left( \widetilde{m}_{i2} - \widetilde{m}_{i1} \right)+ 1 \right) 
     + \widetilde{m}_{i2} - \widetilde{m}_{i1} + 2 \right) }{ \left(  Z_{ i } + 1 \right)^{2} } \\ \\\qquad  
     - \frac{1}{27} \frac{  \left( Z_{i} \left( 2   \left( \widetilde{m}_{i2} - \widetilde{m}_{i1}\right) 
     + 1 \right) + \widetilde{m}_{i2} -\widetilde{m}_{i1} + 2 \right)^{3} }{ \left( Z_{ i } 
     + 1 \right)^{ 3 } } ,
    \end{array}
   \]  
  The allowed values  for the parameters $\delta_{i}$ are in the following range 
   $ 0 < \delta_{ i } < 1 - \widetilde{m}_{ i1 }$.  
   
   Now, the entries in the real orthogonal matrix ${\bf O}$ that
   diagonalize the matrix ${ \bf \bar{M} }_{ i }$, may be expressed as
   \begin{equation}\label{M_ortogonal}
    {\bf O_{i}=}\left(\begin{array}{ccc}
    \left[ \frac{ \widetilde{m}_{i2} f_{i1} }{ {\cal D}_{ i1 } } \right]^{ \frac{1}{2} } & 
    - \left[ \frac{ \widetilde{m}_{i1} f_{i2} }{ {\cal D}_{ i2 } } \right]^{ \frac{1}{2} } & 
    \left[ \frac{ \widetilde{m}_{i1} \widetilde{m}_{i2} \delta_{i} }{ {\cal D}_{i3} } \right]^{ \frac{1}{2} } \\
    \left[ \frac{ \widetilde{m}_{i1} ( 1 - \delta_{i} ) f_{i1} }{ {\cal D}_{i1}  } \right]^{ \frac{1}{2} } & 
    \left[ \frac{ \widetilde{m}_{i2} ( 1 - \delta_{i} ) f_{i2} }{ {\cal D}_{i2} }  \right]^{ \frac{1}{2} }  &  
    \left[ \frac{ ( 1 - \delta_{i} ) \delta_{i} }{ {\cal D}_{i3} } \right]^{ \frac{1}{2} } \\
    - \left[ \frac{ \widetilde{m}_{i1} f_{i2} \delta_{i} }{ {\cal D}_{i1} } \right]^{ \frac{1}{2} } & 
    - \left[ \frac{ \widetilde{m}_{i2} f_{i1} \delta_{i} }{ {\cal D}_{i2} } \right]^{ \frac{1}{2} } &
    \left[ \frac{ f_{i1} f_{i2} }{ {\cal D}_{i3} } \right]^{ \frac{1}{2} }
    \end{array}\right) ,
   \end{equation} 
   where, $ {\cal D}_{i1} = ( 1 - \delta_{i} )( \widetilde{m}_{i1} + \widetilde{m}_{i2} ) ( 1 - \widetilde{m}_{i1} )$, 
   ${\cal D}_{i2} = ( 1 - \delta_{i} )( \widetilde{m}_{i1} + \widetilde{m}_{i2} ) ( 1 + \widetilde{m}_{i2} )$, and 
   ${\cal D}_{i3} = ( 1 - \delta_{i} )( 1 - \widetilde{m}_{i1} )( 1 + \widetilde{m}_{i2} )$.

  \subsection{Mixing Matrices as Functions of the Fermion Masses}
   The unitary matrices ${ \bf U}_{\nu,l}$ occurring in the definition of ${\bf V}_{ _{PMNS} }$, 
   eq.~(\ref{ckm1}), may be written in polar form as $ {\bf U}_{ \nu, l } = {\bf P}_{ \nu, l } {\bf O_{ \nu, l } }$. 
   In this expresion, ${\bf P}_{ \nu, l }$ is the diagonal matrix of phases appearing in the two texutre zeroes mass 
   matrix~(\ref{Polar:FT}). Then, the lepton mixing matrix takes the form 
   \begin{equation}\label{M_unitaria3}
    {\bf V}_{ _{PMNS } }^{ ^{th} } = { \bf O}_{l}^{T} {\bf P}^{ ( \nu - l ) } {\bf O}_{\nu} {\bf K},
   \end{equation}
   where ${ \bf P}^{ ( \nu -l )} = \textrm{diag}\left[1, e^{ i \Phi_{1}  }, e^{ i \Phi_{2} }  \right]$ 
   is the diagonal matrix of the Dirac phases, with $\Phi_{1} =  2\varphi - \phi_{ l }$ and 
   $\Phi_{2} = \varphi - \phi_{l}$. The real orthogonal matrices ${ \bf O}_{ \nu, l } $ are defined in 
   eq.~(\ref{M_ortogonal}).  Thus, the mixing matrix~${\bf V}_{_{PMNS}}^{ ^{th} }$ whose entries are explicit function 
   of the masses of the leptons has the following form~\cite{Barranco:2010we}:
   \begin{equation}
    {\bf V}_{_{ PMNS } }^{ ^{th} } = 
    \left(  \begin{array}{ccc}
     V_{ e 1 }^{ ^{th} } & V_{ e 2 }^{ ^{th} } e^{ i \beta_{ 1 } } &  V_{ e 3 }^{ ^{th} } e^{ i \beta_{ 2 } } \\
     V_{ \mu 1 }^{ ^{th} } & V_{ \mu 2 }^{ ^{th} } e^{ i \beta_{ 1 } } &  V_{ \mu 3 }^{ ^{th} } 
      e^{ i \beta_{ 2 } } \\
     V_{ \tau 1 }^{ ^{th} } & V_{ \tau 2 }^{ ^{th} } e^{ i \beta_{ 1 } } & V_{ \tau 3 }^{ ^{th} } 
     e^{ i \beta_{ 2 } } 
    \end{array} \right),
   \end{equation}
   where
   \begin{equation}\label{elem:pmns}
    \begin{array}{l}
     V_{e1}^{ ^{th} } = 
      \sqrt{ \frac{\widetilde{m}_{\mu} \widetilde{m}_{\nu_{2}} f_{ l1 } f_{ \nu1 } }{ 
      {\cal D}_{ l1 } {\cal D}_{ \nu1 } } } + \sqrt{ \frac{ \widetilde{m}_{e} 
      \widetilde{m}_{\nu_{1}} }{ {\cal D}_{ l1 } {\cal D}_{ \nu 1 } } } 
      \left( \sqrt{ ( 1 - \delta_{l} )( 1 - \delta_{\nu} ) f_{ l1 } f_{ \nu1 } } 
      e^{ i \Phi_{1} } + \sqrt{ \delta_{l} \delta_{\nu} f_{ l2 } f_{ \nu2 } } e^{ i \Phi_{2} } \right), \\
    V_{e2}^{ ^{th} } = 
     - \sqrt{ \frac{\widetilde{m}_{\mu} \widetilde{m}_{\nu_{1}} f_{ l1 } f_{ \nu2 } }{ 
      {\cal D}_{ l1 } {\cal D}_{ \nu2 } } } + \sqrt{ \frac{ \widetilde{m}_{e} 
     \widetilde{m}_{\nu_{2}} }{ {\cal D}_{ l1 } {\cal D}_{ \nu2 } } } \left( \sqrt{ ( 1 - 
     \delta_{l} )( 1 - \delta_{\nu} ) f_{ l1 }  f_{ \nu2 } } e^{ i \Phi_{ 1 } } +
     \sqrt{ \delta_{l} \delta_{\nu} f_{ l2 }  f_{ \nu1 } } e^{ i \Phi_{2} } \right) , \\
    V_{e3}^{ ^{th} }  = 
     \sqrt{ \frac{ \widetilde{m}_{\mu} \widetilde{m}_{\nu_{1}} \widetilde{m}_{\nu_{2}}   
     \delta_{\nu} f_{ l1 } }{ {\cal D}_{ l1 } {\cal D}_{ \nu3} } } + \sqrt{ \frac{ 
     \widetilde{m}_{e} }{ {\cal D}_{ l1 } {\cal D}_{ \nu3 } } } \left( \sqrt{ 
     \delta_{\nu} ( 1 - \delta_{l} ) ( 1 - \delta_{\nu} ) f_{ l1 } } e^{ i \Phi_{1} }
     - \sqrt{ \delta_{e}  f_{ l2 }  f_{ \nu1 }  f_{ \nu2 } }  e^{ i  \Phi_{2}  }\right) , \\
    V_{ \mu1 }^{ ^{th} } = 
     -\sqrt{ \frac{ \widetilde{m}_{e} \widetilde{m}_{\nu_{2}} f_{ l2 } f_{ \nu1 } }{ 
     {\cal D}_{ l2 } {\cal D}_{ \nu1 } } } + \sqrt{ \frac{ \widetilde{m}_{\mu} 
     \widetilde{m}_{\nu_{1}} }{ {\cal D}_{ l2 } {\cal D}_{ \nu1 } } } \left( \sqrt{ ( 1 - 
     \delta_{l} )( 1 - \delta_{\nu} ) f_{ l2 } f_{ \nu1 } } e^{ i \Phi_{1} } + \sqrt{ 
     \delta_{l} \delta_{\nu} f_{ l1 }  f_{ \nu2 } } e^{ i \Phi_{2} } \right) ,\\
    V_{ \mu2 }^{ ^{th} } = 
     \sqrt{ \frac{ \widetilde{m}_{e} \widetilde{m}_{\nu_{1} } f_{ l2 } f_{ \nu2 } }{ 
     {\cal D}_{ l2 } {\cal D}_{\nu 2} } } + \sqrt{ \frac{ \widetilde{m}_{\mu} 
     \widetilde{m}_{\nu_{2}} }{ {\cal D}_{ l2 } {\cal D}_{ \nu2 } } } \left( \sqrt{ ( 1 -
     \delta_{l} ) ( 1 - \delta_{\nu} ) f_{ l2 } f_{ \nu2 } } e^{ i \Phi_{1} } + \sqrt{
     \delta_{l} \delta_{\nu} f_{ l1 } f_{ \nu1 } } e^{ i \Phi_{2}  } \right ) , \\
    V_{ \mu3 }^{ ^{th} } = 
     -\sqrt{ \frac{\widetilde{m}_{e} \widetilde{m}_{\nu_{1}} \widetilde{m}_{\nu_{2}} 
     \delta_{\nu} f_{ l2 } }{ {\cal D}_{ l2 } {\cal D}_{ \nu3 } } } + \sqrt{ 
     \frac{ \widetilde{m}_{\mu} }{ {\cal D}_{ l2 } {\cal D}_{ \nu3 } } } \left( \sqrt{ 
     \delta_{\nu} ( 1 - \delta_{l} ) ( 1 - \delta_{\nu} ) f_{ l2 } } e^{ i \Phi_{1} }
     - \sqrt{ \delta_{l} f_{ l1 } f_{ \nu1 } f_{ \nu2 } } e^{ i \Phi_{2}  } \right ), \\
    V_{ \tau1 }^{ ^{th} } = 
     \sqrt{ \frac{ \widetilde{m}_{e} \widetilde{m}_{\mu} \widetilde{m}_{\nu_{2}} \delta_{l} 
     f_{ \nu1 } }{ {\cal D}_{ l3 } {\cal D}_{ \nu1 } } } + \sqrt{ \frac{ \widetilde{m}_{\nu_{1}} 
     }{ {\cal D}_{ l3 } {\cal D}_{ \nu1 } } } \left( \sqrt{ \delta_{l} ( 1 - \delta_{l} )( 1 - 
     \delta_{\nu} ) f_{ \nu1 } } e^{ i \Phi_{1} } - \sqrt{ \delta_{ \nu } f_{ l1 } f_{ l2 }  
     f_{ \nu2 } } e^{ i \Phi_{2} } \right), \\
    V_{ \tau2 }^{ ^{th} } =  
     - \sqrt{ \frac{ \widetilde{m}_{e} \widetilde{m}_{\mu} \widetilde{m}_{\nu_{1}} \delta_{l} 
     f_{ \nu2 } }{ {\cal D}_{ l3 } {\cal D}_{\nu 2} } } + \sqrt{ \frac{ \widetilde{m}_{\nu_{2}} 
     }{ {\cal D}_{ l3 } {\cal D}_{ \nu2 } } } \left( \sqrt{ \delta_{l} ( 1 - \delta_{l} )( 1 - 
     \delta_{\nu} ) f_{ \nu2 } } e^{ i \Phi_{1} } - \sqrt{ \delta_{\nu} f_{ l1 } f_{ l2 } 
     f_{ \nu1 } } e^{ i \Phi_{2}  } \right ), \\ 
    V_{ \tau3 }^{ ^{th} } = 
     \sqrt{ \frac{ \widetilde{m}_{e} \widetilde{m}_{\mu} \widetilde{m}_{\nu_{1}} 
     \widetilde{m}_{\nu_{2}} \delta_{l} \delta_{\nu} }{ {\cal D}_{ l3 } {\cal D}_{ \nu3 } } } + 
     \sqrt{ \frac{ \delta_{l} \delta_{\nu}( 1 - \delta_{l} ) ( 1 - \delta_{\nu} ) }{ 
     {\cal D}_{ l3 } {\cal D}_{ \nu3 } } } e^{ i \Phi_{1} } + \sqrt{ \frac{  f_{l1 } f_{ l2 }  
     f_{ \nu1 }  f_{ \nu2 } }{ {\cal D}_{ l3 } {\cal D}_{ \nu3 } } } e^{ i \Phi_{2}  } ,
    \end{array}
   \end{equation}
   in these expresions the $f_{\nu}$'s, and ${\cal D}_{\nu}$'s are
   \begin{equation}\label{MsFsDs:leptones}
    \begin{array}{l}
     f_{ \nu(l)1 } = \left( 1 - \widetilde{m}_{\nu_{1}(e)} - \delta_{\nu(l)}  \right), \quad
     {\cal D}_{\nu(l)1} = ( 1 - \delta_{\nu(l)} )( \widetilde{m}_{\nu_{1}(e)} + 
     \widetilde{m}_{\nu_{2}(\mu)} ) ( 1 - \widetilde{m}_{\nu_{1}(e)} ), \\\\
     f_{ \nu(l)2 } = \left( 1 + \widetilde{m}_{\nu_{2}(\mu)} - \delta_{\nu(l)}  \right), \quad
     {\cal D}_{ \nu(l)2} = ( 1 - \delta_{\nu(l)} )( \widetilde{m}_{\nu_{1}(e)} + 
     \widetilde{m}_{\nu_{2}(\mu)} ) ( 1 + \widetilde{m}_{\nu_{2}(\mu)} ),  \\\\
     {\cal D}_{\nu(l)3} = ( 1 - \delta_{\nu(l)} )( 1 - \widetilde{m}_{\nu_{1}(e)} )( 1 + 
     \widetilde{m}_{\nu_{2}(\mu)} ),
    \end{array}
   \end{equation} 
   where $\widetilde{m}_{\nu_{1}(e)} = \frac{ m_{\nu_{1}(e)} }{ m_{\nu_{3}(\tau)} }$, and
   $\widetilde{m}_{\nu_{2}(\mu)} = \frac{ \left| m_{\nu_{2}(\mu)} \right| }{ m_{\nu_{3}(\tau)} }$.
   In the quark sector, the elements of the mixing matrix ${\bf V}_{_{CKM}}$ may also be expressed as functions 
   of the quark masses ratios, the resulting expressions are similar to the expressions obtained above for the 
   elements of the matrix ${\bf V}_{_{PMNS}}$, for more details see~\cite{Barranco:2010we}.

  \subsection{ The $\chi^{2}$ fit for the Lepton Mixing Matrix}
  We made a $\chi^{2}$ fit of the theoretical expressions for the
  modulii of the entries of the lepton mixing matrix $| ( {\bf V}_{ _{
      PMNS } }^{ ^{th} } )_{ij} |$ given in eq.~(\ref{elem:pmns}) to
  the values extracted from experiment as given by Gonzalez-Garcia and
  Maltoni~\cite{GonzalezGarcia:2007ib}. The computation was made using
  the following values for the charged lepton
  masses~\cite{Nakamura:2010zzi}: { \small
    \begin{equation}\label{massChL}
    \begin{array}{l}
     m_{e} = 0.51099~\textrm{MeV}, \;\; m_{\mu}= 105.6583~\textrm{MeV}, \;\;    
     m_{\tau}=1776.82~\textrm{MeV}.
    \end{array} 
   \end{equation} }
   We took for the masses of the left-handed Majorana neutrinos a normal hierarchy. This allows us to write the 
   left-handed Majorana neutrino mass ratios in terms of the neutrino squared mass differences and  the neutrino 
   mass $m_{ \nu_{3} }$ in the following form:
    \begin{equation}
    \begin{array}{l}
     \widetilde{m}_{ \nu_{1} } = \sqrt{ 1 - \frac{ \left( \Delta m_{ 32 }^{ 2 } + 
     \Delta m_{ 21 }^{ 2 } \right) }{ m_{ \nu_{3} }^{ 2 } } }, \;
     \widetilde{m}_{ \nu_{2} } = \sqrt{ 1 - \frac{ \Delta m_{ 32 }^{ 2 } }{ m_{ \nu_{3} }^{ 2 } } }.
    \end{array}
   \end{equation} 
   The numerical values of the neutrino squared mass differences were
   obtained from the global analysis of the experimental data on
   neutrino oscillations given in~\cite{GonzalezGarcia:2007ib}, and we left the mass
   $m_{ \nu_{3} }$ as a free parameter of the $\chi^{2}$ fit. Also,
   the parameters $\delta_{e}$, $\delta_{\nu}$, $\Phi_{1}$ and
   $\Phi_{2} $ were left as
   free parameters to be varied. Hence, in this $\chi^{2}$ fit we have four degrees of freedom. \\
   From the best values obtained for $m_{ \nu_{3} }$ and the
   experimental values of $\Delta m_{ 32 }^{ 2 }$ and $\Delta m_{ 21
   }^{ 2 }$, we obtained the following best values for the neutrino
   masses at 1$\sigma$: { \small \begin{equation}\label{X2:Mnus}
    \begin{array}{l}
     m_{\nu_{1}} = \left( 3.22_{-0.39}^{+0.67} \right) \times 10^{-3}~\textrm{eV}, \,  
     m_{\nu_{2}} = \left( 9.10_{-0.13}^{+0.25} \right) \times 10^{-3}~\textrm{eV}, \, 
     m_{\nu_{3}} = \left( 4.92_{-0.22}^{+0.21} \right) \times 10^{-2}~\textrm{eV}. 
    \end{array}
   \end{equation} }
   The resulting best values of the symmetry breaking parameters are 
   $ \quad$ 
    $\delta_{l} = \left( 6 \pm 2.98 \right) \times 10^{-2}$ and $\quad $ $\delta_{ \nu } = 0.522 _{-0.12}^{+0.09}$,
   and the best values of the Dirac CP violating phases are $\quad$ 
   $ \Phi_{1} = \left( 270 \pm 15 \right)^{\circ}$ and $\Phi_{2} = \left( 180 \pm 10 \right)^{\circ}$. The best values 
   obtained for the modulii of the entries of the $PMNS$ mixing matrix are given in the following expression
   \begin{equation}
    \left| {\bf V}_{ _{PMNS} }^{ ^{th} } \right|_{1\sigma} =
    \left(\begin{array}{ccc}
     0.8204_{-0.010}^{+0.008} & 0.5616_{-0.014}^{+0.012} & 0.1181_{-0.011}^{+0.017}  \\
     0.3748_{-0.031}^{+0.018} & 0.6280_{-0.010}^{+0.019} & 0.6819 \pm 0.025  \\
     0.4345_{-0.020}^{+0.024} & 0.5388_{-0.024}^{+0.022} & 0.7216_{-0.027}^{+0.024}
    \end{array} \right).
   \end{equation} 
   The value of the rephasing invariant related to the Dirac phase is
   $ J_{ l }^{ ^{th} } = \left( 1.8 \pm 0.6 \right) \times 10^{ -2}$.
   In the absence of experimental information about the Majorana
   phases $\beta_{1 }$ and $\beta_{ 2 }$, the two rephasing invariants
   $S_{1}$ and $S_{2}$ associated with the two Majorana phases in the
   ${ \bf V}_{ _{PMNS} }$ matrix, cannot be determined from
   experimental values. Therefore, in order to make a numerical
   estimate of the Majorana phases, we maximized the rephasing
   invariants $S_{1}$ and $S_{2}$, thus obtaining a numerical value
   for the Majorana phases $\beta_{1 }$ and $\beta_{ 2 }$. Then, the
   maximum values of the rephasing invariants are $S_{1}^{ max } =
   -4.9 \times 10^{ -2 }$ and $S_{2}^{ max } = 3.4 \times 10^{ -2 }$,
   with $\beta_{1 }= -1.4^{\circ}$ and $\beta_{ 2 } = 77^{\circ}$. In
   this numerical analysis, the minimum value of the $\chi^{2}$,
   corresponding to the best fit, is $\chi^{2}=0.288$ and the
   resulting value of $\chi^{2}$ for degree of freedom is {\small
     $\frac{\chi^{2}_{min} }{ d.o.f. }=0.075$}. All numerical results
   of the fit are in very good agreement with the values of the moduli
   of the entries in the matrix ${ \bf V}_{ _{PMNS} }$ as given in
   ref.~\cite{GonzalezGarcia:2007ib}.
   
   \subsection{The Mixing Angles}
   The theoretical expressions for the neutrino mixing angles as functions of the charged lepton and 
   neutrino mass  ratios are obtained from eqs.~(\ref{angulosMezclas}), when the theoretical expressions 
   for the modulii of the entries in the $PMNS$ mixing matrix are substituted for 
   $\left| {\bf V}_{ ij } \right| $ in the right hand side of 
   eqs.~(\ref{angulosMezclas}). If we keep only terms at leading order, we obtain:
   \begin{equation}\label{S12L}
    \begin{array}{l}
     \sin^{2}{\theta_{12}^{ l^{th} }} \approx 
      \frac{ f_{ \nu2 } \left \{ \frac{ \widetilde{m}_{\nu_{1}} }{ \widetilde{m}_{\nu_{2}} }  
      + \frac{ \widetilde{m}_{e} }{ \widetilde{m}_{\mu} } \left( 1 - \delta_{ \nu } \right)
      + 2 \sqrt{\frac{ \widetilde{m}_{\nu_{1}} }{ \widetilde{m}_{\nu_{2}} }
      \frac{ \widetilde{m}_{e} }{ \widetilde{m}_{\mu} } \left( 1 - \delta_{ \nu } \right) } \cos{ \Phi_{ _1 } }  
       \right \}}{ \left( 1 + \widetilde{m}_{\nu_{2}} \right) \left( 1 - \delta_{ \nu }  \right) \left( 1 + 
       \frac{\widetilde{m}_{\nu_{1}} }{ \widetilde{m}_{\nu_{2}} } \right)  \left( 1 + 
       \frac{ \widetilde{m}_{e} }{ \widetilde{m}_{\mu} } \right)}  , 
    \end{array}
   \end{equation}
   \begin{equation}\label{S23L}
    \sin^{ 2 } \theta_{23}^{ l^{th} } \approx 
    \frac{ \delta_{ \nu } + \delta_{ e } f_{ \nu2 } - \sqrt{ \delta_{ \nu } \delta_{ e } f_{ \nu2 } } 
    \cos \left( \Phi_{ _1 } - \Phi_{ _2 } \right) }{ \left( 1 + \frac{ \widetilde{m}_{e} }{ \widetilde{m}_{\mu} } 
    \right) \left( 1 + \widetilde{m}_{\nu_{2} } \right) },   
   \end{equation}
   \begin{equation}\label{S13L}
    \begin{array}{l}
     \sin^{ 2 } \theta_{13}^{ l^{th} } \approx  
     \frac{ \delta_{ \nu } \left \{ \frac{ \widetilde{m}_{e} }{
     \widetilde{m}_{\mu} } + \frac{ \widetilde{m}_{\nu_{1} } \widetilde{m}_{\nu_{2}} }{ 
     \left( 1 - \delta_{ \nu } \right)} - 2 \sqrt{ \frac{ \widetilde{m}_{e} }{
     \widetilde{m}_{\mu} } \frac{ \widetilde{m}_{\nu_{1}} \widetilde{m}_{ \nu_{2} } }{ 
     \left( 1 - \delta_{ \nu } \right) } } \cos \Phi_{ _1 } \right \}}{ \left( 1 + 
     \frac{ \widetilde{m}_{e} }{ \widetilde{m}_{\mu} } \right) \left( 1 + \widetilde{m}_{\nu_{2} } \right) } .   
    \end{array}
   \end{equation}
   From eqs.~(\ref{MsFsDs:leptones}) we have that $f_{ \nu2 } = 1 +
   \widetilde{m}_{\nu_{2}} - \delta_{ \nu }$.  The expressions quoted
   above are written in terms of the ratios of the lepton masses,
   defined in eq.~(\ref{T_fritzsch:ev}). When
   the well known values of the charged lepton masses, the values of
   the neutrino masses, eq.~(\ref{X2:Mnus}), the values of the delta
   parameters and the Dirac CP violating phases obtained from the $\chi^2$
   fit in the lepton sector, are inserted in
   eqs.~(\ref{S12L})-(\ref{S13L}), we obtain the following numerical
   values for the mixing angles
   \begin{equation}
    \theta_{12}^{l^{th}} = \left( 34.43_{-0.98}^{+0.85} \right)^{\circ}, \quad  
    \theta_{23}^{l^{th}} = \left( 43.60_{-2.22}^{+1.97} \right)^{\circ}, \quad 
    \theta_{13}^{l^{th}} = \left(  6.80_{-0.66}^{+0.95} \right)^{\circ},
   \end{equation}
   which are in very good agreement with the latest experimental 
   data~\cite{Schwetz:2011zk,PhysRevLett.107.041801,PhysRevLett.107.181802,Abe:2011fz,GonzalezGarcia:2007ib, 
   GonzalezGarcia:2010er,springerlink:10.1134/S106377961104006X,Schwetz:2011qt}. We may conclude 
   that:
   \begin{enumerate}
   \item The strong mass hierarchy of the Dirac fermions 
     produces small and very small mass ratios of charged leptons. 
     Then, the Dirac fermions mass hierarchy is reflected in a similar hierarchy of small and 
     very small Dirac fermion flavour mixing angles~\cite{Barranco:2010we}.
   \item The normal seesaw mechanism type~I which gives very small
     masses to the left-handed Majorana neutrinos with relatively
     large values of the neutrino mass ratio $m_{\nu_1}/m_{\nu_{2}}$
     allows for large $\theta_{12}^{l}$ and $\theta_{23}^{l}$ mixing
     angles (see eqs.~(\ref{S12L})-(\ref{S13L})) .
   \end{enumerate}
   In the quark sector, the mixing angles may also be expressed as
   functions of the quark masses ratios, the resulting expressions are
   similar to the expressions obtained above for the lepton mixing
   angles, for more details see~\cite{Barranco:2010we}.

   \subsection{The effective Majorana masses}
   The theoretical expression for the squared magnitude of the
   effective Majorana neutrino mass of the electron neutrino is:
   {\small \begin{equation}\label{eff-ele}
    \begin{array}{l}
     \left| \langle m_{ee} \rangle^{th} \right|^{2} = 
     m_{\nu_1}^{2}  \left| V_{e1} \right|^{4} + m_{\nu_2}^{2}  \left| V_{e2} \right|^{4} 
     + m_{\nu_3}^{2}  \left| V_{e3} \right|^{4} 
     + 2m_{\nu_1}m_{\nu_2} \left| V_{e1} \right|^{2} \left| V_{e2} \right|^{2} 
     \cos \left( w_{ e1 } - w_{ e2 }\right) \\
     \qquad\qquad \qquad+ 2m_{\nu_1}m_{\nu_3} \left| V_{e1} \right|^{2} \left| V_{e3} \right|^{2} 
     \cos \left( w_{ e1 } - w_{ e3 }\right) +  2m_{\nu_2}m_{\nu_3} \left| V_{e2} \right|^{2} 
     \left| V_{e3} \right|^{2} 
     \cos \left( w_{ e2 } - w_{ e3 }\right), \,
   \end{array}
 \end{equation} }
where $w_{ ei } = \arg \left \{ V_{ei} \right \}  $.
In a similar way, the theoretical expression for the squared magnitude of the effective Majorana neutrino mass of the  muon neutrino is:
{\small \begin{equation}\label{eff-mu}
    \begin{array}{l}
     \left| \langle m_{\mu \mu} \rangle^{th}  \right|^{2} = 
     m_{\nu_1}^{2}  \left| V_{e1} \right|^{4} + m_{\nu_2}^{2}  \left| V_{e2} \right|^{4} 
     + m_{\nu_3}^{2}  \left| V_{e3} \right|^{4} 
     + 2m_{\nu_1}m_{\nu_2} \left| V_{e1} \right|^{2} \left| V_{e2} \right|^{2} 
     \cos \left( w_{ e1 } - w_{ e2 }\right) \\
     \qquad\qquad \qquad+ 2m_{\nu_1}m_{\nu_3} \left| V_{e1} \right|^{2} \left| V_{e3} \right|^{2} 
     \cos \left( w_{ e1 } - w_{ e3 }\right) +  2m_{\nu_2}m_{\nu_3} \left| V_{e2} \right|^{2} \left| V_{e3} \right|^{2} 
     \cos \left( w_{ e2 } - w_{ e3 }\right), 
    \end{array} 
  \end{equation} }
where  $w_{ \mu i } = \arg \left \{ V_{\mu i} \right \}  $.
Expressions for the squared magnitude of effective Majorana masses
$\left| \langle m_{ee} \rangle^{th} \right|^{2}$ and $ \left| \langle
  m_{\mu \mu} \rangle^{th}  \right|^{2}$  with only terms of leading
order are given in Barranco $et$ $al$~\cite{Barranco:2010we}. From
these expressions and the numerical values of the neutrinos masses
given in eq.~(\ref{X2:Mnus}), we obtain the following expressions for
the effective Majorana masses with the phases as free parameters:
{\small \begin{equation}
    \begin{array}{l} 
     \left| \langle m_{ee} \rangle^{th}  \right|^{2} \approx 
     \left \{ 9.41 + 8.29 \cos ( 1^{o} - 2\beta_{1} ) + 4.3  \cos ( 1^{o} - 2w_{e3} ) 
     + 4.31 \cos 2( \beta_{1} - w_{e3} )  \right \} \times 10^{-6}~\textrm{eV}^2,
    \end{array}
   \end{equation} }
   where 
   {\small $ w_{e3} = \arctan \left \{ \frac{0.15 \tan \beta_{2} - 0.013 }{ 0.15 + 0.013 \tan \beta_{2} } 
   \right \} $}. 
   Similarly,
   \begin{equation}
    \begin{array}{l} 
     \left| \langle m_{\mu \mu} \rangle^{th}  \right|^{2} \approx \left \{ 4.8 + 0.17 \cos 2( 44^{o} - w_{\mu 2}   ) 
     + 1.8  \cos 2( w_{\mu 2} - w_{\mu 3} ) \right \} \times 10^{-4}~\textrm{eV}^2,
    \end{array}
   \end{equation}
   where 
   {\small $w_{ \mu 2} \approx \arctan \left \{ \frac{ 0.65 \tan \beta_{1} + 0.13 }{ 0.65 - 0.13 \tan 
   \beta_{1} } \right \} $} and {\small $w_{ \mu 3 } \approx \arctan 
     \left \{  \frac{  \tan \beta_{2} - 0.13  }{ 1 + 0.13 \tan \beta_{2} } \right \} $}. 
     
     In order to make a numerical  estimate of the effective Majorana neutrinos masses 
   $\left| \langle m_{ee} \rangle \right|$ and $\left| \langle m_{\mu \mu} \rangle \right|$,
   we used the following values for the Majorana phases $\beta_{1 }=-1.4^{o}$ and 
   $\beta_{ 2 }=77^{o}$ obtained by maximizing the rephasing invariants $S_{1}$ and $S_{2}$. 
   Then, the numerical value of the effective Majorana neutrino masses are:
   $
    \left| \langle m_{ee} \rangle^{th}  \right| \approx 4.6  \times 10^{ -3 }~\textrm{eV} ,  \quad 
    \left| \langle m_{\mu \mu} \rangle^{th}  \right| \approx 2.1  \times 10^{ -2 }~\textrm{eV} .
   $
   The numerical value of $\left| \langle m_{ee}^{th}  \rangle \right|$ obtained in this theoretical scheme is at
   least two orders of magnitude smaller than the current experimental upper bound. Therefore, it will be very 
   difficult to determine with good precision the numerical value of the effective mass of Majorana neutrinos, 
   since, at present, this value could be of the same order  magnitude as the error bars of the nuclear matrix 
   elements. 
   
  \section{A Minimal invariant Extension of the Standard Model with three Higgs bosons}
  In the Standard Model analogous fermions in different generations
  have identical couplings to all gauge bosons of the strong, weak and
  electromagnetic interactions. Prior to electroweak symmetry breaking, 
  the Lagrangian is chiral and invariant with
  respect to permutations of the left and right fermionic fields, that
  is, it is invariant under the action of the group of permutations
  acting on the flavour indices of the matter fields.

   Since the Standard Model has only one Higgs $SU(2)_{L}$ doublet,
   which can only be an $S_{3}$ singlet, it can only give mass to the
   quark or charged leptons in the $S_{3}$ singlet representation, one
   in each family, without breaking the $S_{3}$ symmetry. Hence, in
   order to impose $S_{3}$ as a fundamental symmetry, unbroken at the
   Fermi scale, we are led to extend the Higgs sector of the theory \cite{Kubo:2003iw}.
   The quark, lepton and Higgs fields are $ Q^T=(u_L,d_L)$, $u_R$, $d_R$,
   $L^T=(\nu_L,e_L)$, $e_R$, $\nu_R$, and $H$, in an
   obvious notation. All these fields have three species, and we
   assume that each one forms a reducible representation ${\bf
     1}_S\oplus{\bf 2}$ of the $S_3$ group.  The doublets carry capital indices $I$ and
   $J$, which run from $1$ to $2$ and the singlets are denoted by
   $Q_3,~u_{3R},~d_{3R},~L_3,~e_{3R},~\nu_{3R}$ and $~H_S$. Note that
   the subscript $3$ denotes the singlet representation and not the
   third generation. The most general renormalizable Yukawa
   interactions for the lepton sector of this model are given by $ {\cal
     L}_Y = {\cal L}_{Y_E}+{\cal L}_{Y_\nu}$, where
   \begin{equation}\label{lage}
    \begin{array}{lll}
     {\cal L}_{Y_E} &=& -Y^e_1\overline{ L}_I H_S e_{IR} - Y^e_3 \overline{ L}_3 H_S e_{3R} 
     - Y^{e}_{2}[~ \overline{ L}_{I}\kappa_{IJ}H_1  e_{JR} + \overline{ L}_{I} \eta_{IJ} H_2  e_{JR}~]\\
     &  & -Y^e_{4}\overline{ L}_3 H_I e_{IR} - Y^e_{5} \overline{ L}_I H_I e_{3R} +~\mbox{h.c.},
    \end{array}
   \end{equation}
   \begin{equation}\label{lagnu}
    \begin{array}{lcl}
     {\cal L}_{Y_\nu} &=& -Y^{\nu}_1\overline{ L}_I (i \sigma_2)H_S^* \nu_{IR} 
     -Y^\nu_3 \overline{ L}_3(i \sigma_2) H_S^* \nu_{3R} -Y^\nu_{4}\overline{ L}_3(i \sigma_2) 
     H_I^* \nu_{IR} \\
     &  &   -Y^{\nu}_{2}[~\overline{ L}_{I}\kappa_{IJ}(i \sigma_2)H_1^*  \nu_{JR}
     + \overline{ L}_{I} \eta_{IJ}(i \sigma_2) H_2^*  \nu_{JR}~]
     -Y^\nu_{5} \overline{ L}_I (i \sigma_2)H_I^* \nu_{3R}+~\mbox{h.c.},
    \end{array}
   \end{equation}
   and
   \begin{equation}\label{kappa}
    \kappa = 
    \left( \begin{array}{cc}
     0 & 1 \\ 
     1 & 0 \\
    \end{array} \right) 
    ~~\mbox{and}~~
    \eta = 
     \left( \begin{array}{cc}
     1 & 0 \\
     0 & -1 \\
    \end{array} \right).
   \end{equation} 
   Furthermore, we add to the Lagrangian the Majorana mass terms for
   the right-handed neutrinos, 
   \begin{equation}
    {\cal L}_{_M} = - \nu_{ R }^T C {\bf M}_{\nu_{R}} \nu_{ R },
   \end{equation}   
   where $C$ is the charge conjugation matrix and 
   ${\bf M}_{\nu_{R}} = \textrm{diag} \left\{ M_1 , M_2, M_3 \right\} $  is the mass matrix for 
   the right-handed
   neutrinos. In the following we will consider two cases: $i$) when
   the first two masses are degenerate, 
   $ M_1 = M_2 \neq M_3$, $ii$) when all masses are different. 

   Due to the presence of three Higgs fields, the Higgs potential of
   the $S_{3}$ invariant extension of the Standard Model is more
   complicated than that of the Standard
   Model~\cite{Kubo:2005sr,EmmanuelCosta:2007zz,Teshima:2012cg}.  This
   potential was first analyzed by Pakvasa and
   Sugawara~\cite{Pakvasa:1977in} who found that in addition to the
   $S_{3}$ symmetry, it has an accidental permutational symmetry
   $S_{2}^{\prime}$: $H_{1}\leftrightarrow H_{2}$, which is not a
   subgroup of the flavour group $S_{3}$. In this communication, we
   will assume that the vacuum respects the accidental
   $S_{2}^{\prime}$ symmetry of the Higgs potential and that $\langle
   H_{1} \rangle = \langle H_{2} \rangle. $ With these assumptions,
   the Yukawa interactions, eqs.~(\ref{lage})-(\ref{lagnu}) yield mass
   matrices, for all fermions in the theory, of the general
   form~\cite{Kubo:2003iw}
   \begin{equation}\label{matGene}
    {\bf M} = 
    \left( \begin{array}{ccc}
     \mu_{1} +\mu_{2} & \mu_{2} & \mu_{5} \\ 
     \mu_{2} & \mu_{1}-\mu_{2} & \mu_{5}   \\ 
     \mu_{4} & \mu_{4}&  \mu_{3}
    \end{array}\right). 
   \end{equation}
   
   In principle, all entries in the mass matrices can be complex since
   there is no restriction coming from the flavour symmetry
   $S_{3}$. Therefore, there are $4 \times 5=20$ complex parameters in
   the mass matrices, which should be compared with $4 \times 9=36 $
   of the SM, including the Majorana masses of the left-handed
   neutrinos.
   
   \subsection{The mass matrices in the leptonic sector and $Z_{2}$ symmetry}
   A further reduction of the number of parameters in the leptonic
   sector may be achieved by means of an Abelian $Z_{2}$ symmetry
   \cite{Kubo:2003iw}. A possible set of charge assignments of
   $Z_{2}$, compatible with the experimental data on masses and
   mixings in the leptonic sector is given in Table~\ref{table1}.
    \begin{table}
    \begin{center}
     \caption{$Z_2$ assignment in the leptonic sector.}\label{table1}    
      \begin{tabular}{|c|c|} \hline
       $-$ &  $+$ \\ \hline
       $H_S, ~\nu_{3R}$ & $H_I, ~L_3, ~L_I, ~e_{3R},~ e_{IR},~\nu_{IR}$ \\ \hline
      \end{tabular}
     \end{center} 
    \end{table}
   These $Z_2$ assignments forbid the following Yukawa couplings 
   $ Y^e_{1} = Y^e_{3}= Y^{\nu}_{1}= Y^{\nu}_{5}=0$.
   Therefore, the corresponding entries in the mass matrices vanish, {\it i.e.}, 
   $\mu_{1}^{e}=\mu_{3}^{e}=0$ and 
   $\mu_{1}^{\nu}=\mu_{5}^{\nu}=0$. 

   The mass matrix of the charged leptons takes the form
   \begin{equation}\label{charged-leptons-m}
    { \bf M}_{e} = m_{\tau}
     \left( \begin{array}{ccc}
     \tilde{\mu}_{2} &  \tilde{\mu}_{2} & \tilde{\mu}_{5} \\  
     \tilde{\mu}_{2} & -\tilde{\mu}_{2} & \tilde{\mu}_{5}  \\ 
     \tilde{\mu}_{4} &  \tilde{\mu}_{4} & 0
    \end{array}\right).
   \end{equation}
   The resulting expression for ${ \bf M}_e$, reparametrized in terms of its eigenvalues and written
    to order 
   $\left(m_{\mu}m_{e}/m_{\tau}^{2}\right)^{2}$ and $x^{4}=\left(m_{e}/m_{\mu}\right)^4$, is
   \begin{equation}\label{emass}
    { \bf M}_{e}\approx m_{\tau} \left( 
    \begin{array}{ccc}
    \frac{1}{\sqrt{2}}\frac{\tilde{m}_{\mu}}{\sqrt{1+x^2}} & \frac{1}{\sqrt{2}}\frac{\tilde{m}_{\mu}}{\sqrt{1+x^2}}&   
    \frac{1}{\sqrt{2}} \sqrt{\frac{1+x^2-\tilde{m}_{\mu}^2}{1+x^2}} \\ \\
    \frac{1}{\sqrt{2}}\frac{\tilde{m}_{\mu}}{\sqrt{1+x^2}} &-\frac{1}{\sqrt{2}}\frac{\tilde{m}_{\mu}}{\sqrt{1+x^2}}& 
    \frac{1}{\sqrt{2}} \sqrt{\frac{1+x^2-\tilde{m}_{\mu}^2}{1+x^2}} \\ \\
    \frac{\tilde{m}_{e}(1+x^2)}{\sqrt{1+x^2-\tilde{m}_{\mu}^2}}e^{i\delta_{e}} & 
    \frac{\tilde{m}_{e}(1+x^2)}{\sqrt{1+x^2-\tilde{m}_{\mu}^2}}e^{i\delta_{e}} & 0
    \end{array} \right). 
   \end{equation}
   This approximation is numerically exact up to order $10^{-9}$ in
   units of the $\tau$ mass. Notice that this matrix has no free
   parameters other than the Dirac phase $\delta_e$.

   The unitary matrix ${ \bf U}_{eL}$ that enters in the definition of
   the mixing matrix, ${ \bf V}_{_{PMNS}}$, is calculated from ${ \bf
     U}_{eL}^{\dag} { \bf M}_{e} { \bf M}_{e}^{\dag}{ \bf U}_{eL}
   =\mbox{diag} \left(m_{e}^{2},m_{\mu}^{2},m_{\tau}^{2} \right)$,
   where $m_{e}$, $m_{\mu}$ and $m_{\tau}$ are the masses of the
   charged leptons, written to the same order of magnitude as ${\bf
     M}_e$ is ${ \bf U}_{eL} = {\bf P}^{e} {\bf O}^{e}$, where ${\bf
     P}^{e} = \textrm{diag} \left\{ 1, 1, e^{i \delta_{e}} \right\}$
   and {\small \begin{equation}\label{unitary-leptons-2} {\bf O}^{e}
       \approx \left( \begin{array}{ccc} \frac{1}{\sqrt{2}} x \frac{ (
             1 + 2 \tilde{m}_{\mu}^2 + 4x^2 + \tilde{m}_{\mu}^4 +
             2\tilde{m}_{e}^2)}{ \sqrt{ 1 + \tilde{m}_{\mu}^2 + 5x^2
               -\tilde{m}_{\mu}^4 - \tilde{m}_{\mu}^6 +
               \tilde{m}_{e}^2 + 12x^4 } } & - \frac{ 1 }{ \sqrt{2}}
           \frac{(1-2\tilde{m}_{\mu}^2+\tilde{m}_{\mu}^4-2
             \tilde{m}_{e}^2) }{
             \sqrt{1-4\tilde{m}_{\mu}^2+x^2+6\tilde{m}_{\mu}^4-4\tilde{m}_{\mu}^6-5\tilde{m}_{e}^2}}
           & \frac{1}{\sqrt{2}}
           \\ \\
           -\frac{1}{\sqrt{2}}x \frac{(
             1+4x^2-\tilde{m}_{\mu}^4-2\tilde{m}_{e}^2 ) }{
             \sqrt{1+\tilde{m}_{\mu}^2+5x^2-\tilde{m}_{\mu}^4-\tilde{m}_{\mu}^6+\tilde{m}_{e}^2+12x^4}}
           &
           \frac{1}{\sqrt{2}}\frac{(1-2\tilde{m}_{\mu}^2+\tilde{m}_{\mu}^4)
           }{
             \sqrt{1-4\tilde{m}_{\mu}^2+x^2+6\tilde{m}_{\mu}^4-4\tilde{m}_{\mu}^6-5\tilde{m}_{e}^2}}
           &
           \frac{1}{\sqrt{2}} \\ \\
           -
           \frac{\sqrt{1+2x^2-\tilde{m}_{\mu}^2-\tilde{m}_{e}^2}(1+\tilde{m}_{\mu}^2+x^2-2\tilde{m}_{e}^2)
           }{
             \sqrt{1+\tilde{m}_{\mu}^2+5x^2-\tilde{m}_{\mu}^4-\tilde{m}_{\mu}^6+\tilde{m}_{e}^2+12x^4}}
           &
           -x\frac{(1+x^2-\tilde{m}_{\mu}^2-2\tilde{m}_{e}^2)\sqrt{1+2x^2-\tilde{m}_{\mu}^2-\tilde{m}_{e}^2}
           }{
             \sqrt{1-4\tilde{m}_{\mu}^2+x^2+6\tilde{m}_{\mu}^4-4\tilde{m}_{\mu}^6-5\tilde{m}_{e}^2}}
           &
           \frac{\sqrt{1+x^2}\tilde{m}_{e}\tilde{m}_{\mu}}{\sqrt{1+x^2-\tilde{m}_{\mu}^2}}
    \end{array} \right),
   \end{equation} }
   in this expression, as before, $\tilde{m}_{\mu}=m_{\mu}/m_{\tau}$, $\tilde{m}_{e}=m_{e}/m_{\tau}$ 
   and $x=m_{e}/m_{\mu}$~\cite{Felix:2006pn}. 

  \subsubsection{Neutrino masses and mixings with two degenerate masses of the right-handed neutrinos}
  According to the $Z_2$ selection rule the mass matrix of the Dirac
  neutrinos takes the form
  \cite{Kubo:2003iw,Felix:2006pn,Mondragon:2007af}
   \begin{equation}\label{neutrinod-m}
    {\bf M}_{\nu_D} = 
    \left( \begin{array}{ccc}
     \mu^{\nu}_{2} & \mu^{\nu}_{2}  & 0 \\ 
     \mu^{\nu}_{2} & -\mu^{\nu}_{2} & 0 \\ 
     \mu^{\nu}_{4} & \mu^{\nu}_{4}&  \mu^{\nu}_{3}
    \end{array}\right),
   \end{equation}
   and considering the following form to the mass matrix of right-handed neutrinos
   {\small ${\bf M}_{\nu_R} = \textrm{diag} \left \{ M_{1}, M_{1} , M_{3} \right \}$}, 
   the mass matrix for the left-handed 
   Majorana neutrinos, ${\bf M}_{\nu_L}$, obtained from the seesaw mechanism type~I is,
   \begin{equation}\label{m-nu}
    {\bf M}_{\nu_L}  = 
    \left( \begin{array}{ccc}
     2 (\rho^{\nu}_{2})^2 & 0 & 2 \rho^{\nu}_2 \rho^{\nu}_{4} \\ 
     0 & 2 (\rho^{\nu}_{2})^2 & 0 \\ 
     2 \rho^{\nu}_2 \rho^{\nu}_{4} & 0  &  2 (\rho^{\nu}_{4})^2 + (\rho^{\nu}_3)^2
    \end{array}\right),
   \end{equation}
   where $\rho_2^\nu =(\mu^{\nu}_2)/M_{1}^{1/2}$, $\rho_4^\nu
   =(\mu^{\nu}_4)/M_{1}^{1/2}$ and $\rho_3^\nu
   =(\mu^{\nu}_3)/M_{3}^{1/2}$; $M_{1}$ and $M_{3}$ are the masses of
   the right handed neutrinos.  The complex symmetric neutrino mass
   matrix ${ \bf M}_{\nu_L}$ may be brought to a diagonal form by the
   transformation ${ \bf U}_{\nu}^{T} { \bf M}_{\nu} { \bf U}_{\nu} =
   \mbox{diag}\left(|m_{\nu_{1}}|e^{i\phi_{1}},|m_{\nu_{2}}|e^{i\phi_{2}},|m_{\nu_{3}}|e^{i\phi_{\nu}}\right)$,
   where ${ \bf U}_{\nu}$ is the matrix that diagonalizes the matrix $
   { \bf M}_{\nu_L}^{\dagger} { \bf M}_{\nu_L}$ .  This allows us to
   reparametrize the matrices ${ \bf M}_{ \nu_L }$ and ${ \bf U}_\nu$
   in terms of the complex neutrino masses, which have the following
   form:
   \begin{equation}\label{m-nu2}
    { \bf M}_{\nu_L} = 
    \left( \begin{array}{ccc}
     m_{\nu_{3}} & 0 & \sqrt{(m_{\nu_{3}}-m_{\nu_{1}})(m_{\nu_{2}}-m_{\nu_{3}})}e^{-i\delta_{\nu}} \\ 
     0 & m_{\nu_{3}}  & 0 \\
     \sqrt{(m_{\nu_{3}}-m_{\nu_{1}})(m_{\nu_{2}}-m_{\nu_{3}})} e^{-i\delta_{\nu}} & 0  & 
     (m_{\nu_{1}}+m_{\nu_{2}}-m_{\nu_{3}})e^{-2i\delta_{\nu}}
    \end{array}\right)
   \end{equation}
   and ${ \bf U}_{\nu} = { \bf P}_{ \nu } { \bf O}_{\nu}$ where 
   ${ \bf P}_{ \nu } = \textrm{diag} \left \{ 1, 1,  e^{i\delta_{\nu}}  \right \}$ and 
   \begin{equation}\label{ununew}
    { \bf O}_{\nu} =  \left( \begin{array}{ccc}
     \cos \eta & \sin \eta & 0 \\
     0 & 0 & 1 \\
     - \sin \eta  & \cos \eta & 0
    \end{array} \right) =
    \left( \begin{array}{ccc}
     \sqrt{ \frac{ m_{\nu_{2}} - m_{\nu_{3}} }{ m_{\nu_{2}} - m_{\nu_{1}} } } & 
     \sqrt{ \frac{ m_{\nu_{3}} - m_{\nu_{1}} }{ m_{\nu_{2}} - m_{\nu_{1}} } } & 0 \\
     0 & 0 & 1 \\ 
     - \sqrt{ \frac{ m_{\nu_{3} } - m_{\nu_{1}} }{ m_{\nu_{2}} - m_{\nu_{1}} } }  &
     \sqrt{ \frac{m_{\nu_{2} } - m_{\nu_{3}} }{ m_{\nu_{2} } - m_{\nu_{1}} } } & 0
    \end{array} \right).
   \end{equation}
   The unitarity of ${ \bf U}_{\nu}$ constrains $\sin \eta$ to be real and thus 
   $|\sin \eta|\leq 1$, this condition fixes the phases $\phi_{1}$ and $\phi_{2}$ as 
   $|m_{\nu_{1}}|\sin \phi_{1}=|m_{\nu_{2}}|\sin \phi_{2}=|m_{\nu_{3}}|\sin \phi_{\nu}$. 
   The only free parameters in the matrices ${\bf M}_\nu$ and ${ \bf U}_\nu$ are the  phase 
   $\phi_{\nu}$, implicit in $m_{\nu_{1}}$, $m_{\nu_{2}}$ and $m_{\nu_{3}}$, and the Dirac 
   phase $\delta_{\nu}$.

   The neutrino mixing matrix ${\bf V}_{_{PMNS}}$, is the product 
   $ {\bf U}_{eL}^{\dagger} {\bf U}_{\nu} {\bf K}$,   where 
   ${ \bf K} = \textrm{diag} \left(1,e^{i\alpha},e^{i\beta} \right)$ is the diagonal matrix of the 
   Majorana 
   phase factors.  Therefore, the theoretical mixing matrix ${\bf V}_{_{PMNS}}^{^{th}}$, is given by
   \begin{equation}\label{vpmns2}
    {\bf V}_{_{PMNS} }^{^{th}}=
    \left( \begin{array}{ccc}
     O_{11}^{e}\cos \eta + O_{31}^{e} \sin \eta e^{i\delta_{l}} & O_{11}^{e} \sin \eta - O_{31}^{e}  
     \cos \eta e^{i\delta_{l}} & - O_{21}^{e}   \\ \\
    -O_{12}^{e} \cos \eta + O_{32}^{e} \sin \eta e^{i\delta_{l}} & -O_{12}^{e} \sin \eta - O_{32}^{e} 
    \cos \eta e^{i\delta_{l}} & 
    O_{22}^{e}  \\ \\
    O_{13}^{e} \cos \eta - O_{33}^{e} \sin \eta e^{i\delta_{l}} & O_{13}^{e}  \sin \eta + O_{33}^{e} 
    \cos \eta e^{i\delta_{l}} & O_{23}^{e}  
    \end{array} \right)  {\bf K},
   \end{equation}
   where $\cos \eta$ and $\sin \eta$ are given in eq.~(\ref{ununew}), $O^{e}_{ij}$ are given in
   eq.~(\ref{unitary-leptons-2}), and $\delta_{l}=\delta_{\nu}-\delta_{e}$. 

  \begin{flushleft}
   {  \bf The mixing angles }
  \end{flushleft}  
  The magnitudes of the reactor and atmospheric mixing angles,
  $\theta_{13}^{l^{th}}$ and $\theta_{23}^{l^{th}}$, are determined by
  the masses of the charged leptons only. Keeping only terms of order
  $(m_{e}^2/m_{\mu}^2)$ and $(m_{\mu}/m_{\tau})^4$, we get \cite{Mondragon:2007af}
   \begin{equation}
    \begin{array}{lr}
     \sin \theta_{13}^{l^{th}} \approx 
     \frac{1}{\sqrt{2}}x
     \frac{( 1+4x^2-\tilde{m}_{\mu}^4)}{\sqrt{1+\tilde{m}_{\mu}^2+5x^2-\tilde{m}_{\mu}^4}}, &
     \sin \theta_{23}^{l^{th}} \approx  
     \frac{1}{\sqrt{2}}\frac{1+\frac{1}{4}x^2-2\tilde{m}_{\mu}^2+\tilde{m}_{\mu}^4} 
     {\sqrt{1-4\tilde{m}_{\mu}^2+x^2+6\tilde{m}_{\mu}^4}}.
    \end{array}
   \end{equation}
   The magnitude of the solar angle depends on the charged lepton and neutrino masses as well as the Dirac and 
   Majorana phases
   \begin{equation}\label{tan2}
    |\tan \theta_{12}^{l^{th}}|^2= \frac{\displaystyle{m_{\nu_{2}}-m_{\nu_{3}}}}{
    \displaystyle{m_{\nu_{3}}-m_{\nu_{1}}}}\left(\frac{1-2\frac{O_{11}}{O_{31}}\cos   
    \delta_{l} \sqrt{\frac{\displaystyle{m_{\nu_{3}}-m_{\nu_{1}}}}{
    \displaystyle{m_{\nu_{2}}-m_{\nu_{3}}}}}+\left(\frac{O_{11}}{O_{31}}\right)^2\frac{\displaystyle{m_{\nu_{3}}-
    m_{\nu_{1}}}}{\displaystyle{m_{\nu_{2}}-m_{\nu_{3}}}}}{1+2\frac{O_{11}}{O_{31}} \cos 
    \delta_{l}\sqrt{\frac{\displaystyle{m_{\nu_{2}}-m_{\nu_{3}}}}{ \displaystyle{m_{\nu_{3}} 
    -m_{\nu_{1}}}}}+\left(\frac{O_{11}}{O_{31}}\right)^2\frac{\displaystyle{m_{\nu_{2}}-m_{\nu_{3}}}}{
    \displaystyle{m_{\nu_{3}}-m_{\nu_{1}}}} }\right) . 
   \end{equation}
   The dependence of $\tan \theta_{12}^{l^{th}}$ on the Dirac phase
   $\delta_{l}$, see (\ref{tan2}), is very weak, since $O_{31}\sim 1$
   but $O_{11}\sim 1/\sqrt{2}(m_e/m_\mu)$. Hence, we may neglect it
   when comparing (\ref{tan2}) with the data on neutrino mixings. The
   dependence of $\tan \theta_{12}^{l^{th}}$ on the phase $\phi_{\nu}$
   and the physical masses of the neutrinos enters through the ratio
   of the neutrino mass differences, it can be made explicit with the
   help of the unitarity constraint on ${ \bf U}_{\nu}$ as
   \begin{equation}\label{tansq}
    \frac{\displaystyle{m_{\nu_{2}}-m_{\nu_{3}}}}{\displaystyle{m_{\nu_{3}}-m_{\nu_{1}}}}=
    \frac{(|m_{\nu_{2}}|^2-|m_{\nu_{3}}|^{2}\sin^{2}\phi_{\nu})^{1/2}-|m_{\nu_{3}}| |\cos \phi_{\nu}|}
    {(|m_{\nu_{1}}|^{2}-|m_{\nu_{3}}|^{2}\sin^{2}\phi_{\nu})^{1/2}+|m_{\nu_{3}}| |\cos \phi_{\nu}|}.
   \end{equation}
   Similarly, the Majorana phases are given by $\sin 2\alpha=\sin(\phi_{1}-\phi_{2})$ and 
   $\sin 2\beta=\sin(\phi_{1}-\phi_{\nu})$ where
   \begin{equation}
    \begin{array}{l}
     \sin 2\alpha = \frac{|m_{\nu_{3}}|\sin\phi_{\nu}}{|m_{\nu_{1}}||m_{\nu_{2}}|}
     \left(\sqrt{|m_{\nu_{2}}|^2-|m_{\nu_{3}}|^{2}\sin^{2}\phi_{\nu}}+\sqrt{|m_{\nu_{1}}|^{2}
     -|m_{\nu_{3}}|^{2}\sin^{2}\phi_{\nu}}\right), \\ \\
     \sin 2\beta = 
     \frac{\sin\phi_{\nu}}{|m_{\nu_{1}}|}\left(|m_{\nu_{3}}|\sqrt{1-\sin^{2}\phi_{\nu}}
     +\sqrt{|m_{\nu_{1}}|^{2}  -|m_{\nu_{3}}|^{2}\sin^{2}\phi_{\nu}}\right).
    \end{array}
   \end{equation}
   A more complete and detailed discussion of the neutrino mixing
   matrix $V_{PMNS}$ and the Majorana phases obtained
   in this model is given in refs.~\cite{Kubo:2003iw,Felix:2006pn,Mondragon:2007af,Mondragon:2008gm}.\\

   In the present $S_3$-invariant extension of the Standard Model, the
   experimental restriction $|\Delta m_{12}^2|<|\Delta m_{13}^2|$
   implies an inverted neutrino mass spectrum
   $|m_{\nu_{3}}|<|m_{\nu_{1}}|<|m_{\nu_{2}}|$~\cite{Kubo:2003iw,Mondragon:2007af}. In
   this model, the reactor and atmospheric mixing angles,
   $\theta_{13}^{l^{th}}$ and $\theta_{23}^{l^{th}}$, are determined
   by the masses of the charged leptons only as
   \begin{equation}\label{cond1}
    \begin{array}{ll}
     \sin^{2} \theta_{13}^{ l^{th} } = 1.1 \times 10^{-5}, & 
     \sin^{2} \theta_{23}^{ l^{th} } = 0.5.     
    \end{array}
   \end{equation}
   Even in this simplified analysis it is clear that the $S_{3}$
   symmetry gives a non-vanishing reactor mixing angle, within the
   bounds of MINOS~\cite{PhysRevLett.107.181802},  albeit small.  The atmospheric angle is in 
   very good agreement with the recent experimental data. As can be seen from equations
   (\ref{tan2}) and (\ref{tansq}), the solar angle is sensitive to the
   differences of the squared neutrino masses and the phase
   $\phi_{\nu}$ but is only weakly sensitive to the charged lepton
   masses. If the small terms proportional to $O_{11}$ and
   $O_{11}^{2}$ are neglected in (\ref{tan2}), we obtain
   \begin{equation}
    \tan^2 \theta_{12}^{l^{th}} =
    \frac{ ( \Delta m_{12}^{2} + \Delta m_{13}^2 + | m_{ \nu_{3} } |^{2} \cos^2 \phi_\nu  )^{1/2} 
    - | m_{ \nu_{3} } |  \cos \phi_{\nu}  }{ ( \Delta m_{13}^2 + |m_{\nu_{3}| }^{2} \cos^2 \phi_\nu)^{1/2} + |
    m_{\nu_{3}}|   \cos \phi_\nu  }.
   \end{equation}
   From this equation, we may readily derive expressions for the
   neutrino masses in terms of $\tan \theta_{12}^{l^{th}}$, $\cos
   \phi_\nu$ and the differences of the squared neutrino masses
   \begin{equation}
    | m_{ \nu_{3} } | = 
    \frac{ \sqrt{ \Delta m_{13}^2 } }{ 2 \tan \theta_{12}^{l^{th}} \cos \phi_{\nu} }  
    \frac{ 1  - \tan^4 \theta_{12}^{l^{th}} + r^2 }{ \sqrt{ 1 + \tan^2 \theta_{12}^{l^{th}}} \sqrt{ 1 
    + \tan^2 \theta_{12}^{l^{th}}  + r^2 } },
   \end{equation}
   $|m_{\nu_{1}}|= \sqrt{ |m_{\nu_{3}}|^2 + \Delta m_{13}^2}$, and 
   $|m_{\nu_{2}}|= \sqrt{ |m_{\nu_{3}}|^2 + \Delta m_{13}^2(1+r^2) }$, where 
   $r^2=\Delta m_{12}^2/\Delta m_{13}^2\approx 3\times 10^{-2}$.  As $r^2<<1$, 
   the sum of the neutrino masses is
   \begin{eqnarray}\label{MsasS3:1}
    \sum_{i=1}^{3} |m_{\nu_{i}}|\approx\frac{\sqrt{\Delta m_{13}^2}}{2\cos \phi_{\nu}
     \tan \theta_{12}^{l^{th}} }\left(1+2\sqrt{1+2\tan^2 \theta_{12}^{l^{th}} 
     \cos 2 \phi_\nu +\tan^4  \theta_{12}^{l^{th}} } 
     - \tan^2 \theta_{12}^{l^{th}} \right). 
   \end{eqnarray} 
   The most restrictive cosmological upper bound for this sum is
   {\small $\sum |m_{\nu}|\leq 0.17 eV$}~\cite{Seljak:2006bg}.  This
   upper bound and the experimentally determined values of $\tan
   \theta_{12}$ and $\Delta m_{ij}^{2}$, give a lower bound for $\cos
   \phi_{\nu} \geq 0.55$ or $0\leq \phi_\nu \leq 57^{\circ}$. The
   neutrino masses $|m_{\nu_i}|$ assume their minimal values when
   $\cos \phi_\nu=1$. When $\cos \phi_\nu$ takes values in the range
   $0.55 \leq \cos \phi \leq 1$, the neutrino masses change very
   slowly with $\cos \phi_\nu$. In the absence of experimental
   information we will assume that $\phi_\nu$ vanishes. Hence, setting
   $\phi_\nu=0$ in our formula, we find
   \begin{equation}
     m_{\nu_1}= 0.052~eV \quad m_{\nu_2}=0.053 ~eV \quad m_{\nu_3}= 0.019~eV.
   \end{equation}
   The computed sum of the neutrino masses is {\small $ \left(
       \sum_{i=1}^3 |m_{\nu_{i}}|\right)^{th} = 0.13 ~eV$}, which is
   consistent with the cosmological upper bound~\cite{Seljak:2006bg},
   as expected, since we used the cosmological bound to fix the bound
   on $\cos \phi_{\nu}$.
%
   The effective Majorana mass in neutrinoless double beta decay
   $\langle m_{ee} \rangle$, is defined in
   eq.~(\ref{masa_eff.19}). The most stringent bound on $\langle
   m_{ee} \rangle$, obtained from the analysis of the data collected
   by the Heidelberg-Moscow experiment on the process neutrinoless
   double beta decay in enriched
   Ge~\cite{KlapdorKleingrothaus:2000sn}, is $ \langle m_{ee} \rangle
   < 0.3~eV$. In this framework for a preliminary analysis we are
   assuming that the Majorana phases vanish, thus we get
   \begin{equation}
    \langle m_{ee} \rangle ^{th}=0.053 ~eV
   \end{equation}
   well below the experimental upper bound.
   
  \begin{flushleft}
    {\bf Deviation  of the mixing matrix ${\bf V}_{PMNS}^{th}$ from the tri-bimaximal form}
  \end{flushleft} 
 
   The previous results on neutrino masses and mixings weakly depend on the Dirac phase $\delta$, for simplicity we 
   will assume in this work that $\delta=\pi/2$. We may write the mixing matrix 
   as follows, $ {\bf V}_{_{PMNS}}^{th} = {\bf V}_{_{PMNS}}^{tri} +   \Delta {\bf V}_{_{PMNS}}^{tri}$
   where the tri-bimaximal form ${\bf V}_{_{PMNS}}^{tri}$~\cite{Harrison:2002er} is
   \begin{equation}\label{tribimax}
    {\bf V}_{PMNS}^{tri}=\left(\begin{array}{ccc}
     \sqrt{\frac{2}{3}} &\sqrt{\frac{1}{3}} & 0\\
     -\sqrt{\frac{1}{6}} &\sqrt{\frac{1}{3}} & -\sqrt{\frac{1}{2}}\\
     -\sqrt{\frac{1}{6}} &\sqrt{\frac{1}{3}} & \sqrt{\frac{1}{2}}
    \end{array}\right), \; \textrm{and} \;
    \Delta{\bf  V}_{PMNS}^{tri}=\Delta {\bf  V}_{e} 
    + \delta t_{12}\frac{(\sqrt{2}+\delta t_{12})}{g(\delta t_{12})}\Delta{\bf V}_{\nu},
   \end{equation}
   where $g(\delta t_{12})=1+\frac{2}{3}\delta t_{12}(\sqrt{2}+\delta
   t_{12})$. All entries in $\Delta {\bf V}_{e}$ are proportional to
   $\left( m_{e}/m_{\mu} \right)^{2}$ except $\left( \Delta {\bf
       V}_{e} \right)_{13}$ that is proportional to $\left(
     m_{e}/m_{\mu} \right)$.
   
   The value for $\delta t_{12}$, which is a small parameter, fixes
   the scale and the origin of the neutrino mass matrix.  If we take
   for $\delta t_{12}$ the experimental central value $\delta
   t_{12}\approx -0.04$, we obtain $|m_{\nu_{2}}|\approx0.056$~eV,
   $|m_{\nu_{1}}| \approx 0.055$~eV, and $|m_{\nu_{3}}| \approx
   0.022$~eV~\cite{Mondragon:2007af}.  When we take for $\delta
   t_{12}$ the tri-bimaximal value $\delta t_{12}=0$, the neutrino
   masses are $ m_{\nu_1}= 0.0521$~eV, $m_{\nu_2}=0.0528$~eV, and
   $m_{\nu_3}= 0.0178$~eV. For a detailed discussion on this subject 
   see~\cite{Mondragon:2007jx}.
   
   In both cases the $S_{3}$ invariant extension of the SM predicts an
   inverted hierarchy. Since the tri-bimaximal value of $\delta
   t_{12}$ differs from the experimental central value by less than
   $6\%$ of $\tan \theta_{12}$, the difference in the corresponding
   numerical values of the neutrino masses are not significative
   within the present experimental uncertainties.

  \subsubsection{Neutrino masses and mixings when the masses of the right-handed neutrinos are 
  non-degenerate }
   In the minimal $S_{3}$-invariant extension of Standard Model, the Yukawa interactions and the 
   $S_{3} \times Z_{2}$ flavour symmetry yield a mass matrix for the Dirac neutrinos of the 
   form~(\ref{neutrinod-m}). The masses of the left-handed Majorana neutrinos, ${\bf M}_{\nu_L}$, are generated by the  
   seesaw mechanism type~I, eq.~(\ref{subibajadef}), where ${\bf M}_{\nu_R}$ is the mass matrix of the 
   right-handed neutrinos, which we take to be real and diagonal but non-degenerate
   $ {\bf M}_{\nu_R} = \mbox{diag}(M_1,M_2,M_3)$. Then, the mass matrix ${\bf M}_{\nu_L}$ takes the form
   \begin{equation}\label{NeuMajorana}
    {\bf M_{\nu_L}} =
    \left( \begin{array}{ccc}
     \frac{ 2 \left( \mu^{\nu}_{2} \right)^{2} }{ \overline{M} } & 
     \frac{ 2 \lambda \left( \mu^{\nu}_{2} \right)^{2} }{ \overline{M} } & 
     \frac{ 2 \mu^{\nu}_{2} \mu^{\nu}_{4} }{ \overline{M} } \\ \\
     \frac{ 2 \lambda \left( \mu^{\nu}_{2} \right)^{2} }{ \overline{M} } &  
     \frac{ 2 \left( \mu^{\nu}_{2} \right)^{2} }{ \overline{M} } &
     \frac{ 2 \mu^{\nu}_{2} \mu^{\nu}_{4} \lambda }{ \overline{M} } \\ \\ 
     \frac{ 2 \mu^{\nu}_{2} \mu^{\nu}_{4} }{ \overline{M} } & 
     \frac{ 2 \mu^{\nu}_{2} \mu^{\nu}_{4} \lambda }{ \overline{M} } &
     \frac{ 2 \left( \mu^{\nu}_{4} \right)^{2} }{ \overline{M} } + 
     \frac{ \left( \mu^{\nu}_{3} \right)^{2} }{ M_{3} }
    \end{array} \right), \,
    \begin{array}{l}
     \lambda =  \left( \frac{ M_2 - M_1 }{ M_1 + M_2 } \right),
     \, \textrm{and} \,
     \overline{M} = 2\frac{ M_1 M_2 }{ M_2 + M_1  }.
    \end{array}
   \end{equation}
   When the first two right-handed neutrino masses are equal, the
   parameter $\lambda$ vanishes and we recover the expresion for ${\bf
     M}_{\nu_L}$ given in Kubo {\it et al}~\cite{Kubo:2003iw},
   eq.~(\ref{m-nu}) in the present paper, which leads to the results
   presented in the previous
   section~\cite{Felix:2006pn,Mondragon:2008gm,Mondragon:2007jx,Mondragon:2007nk,Mondragon:2007af}.

   Since we assumed the right-handed neutrino mass matrix ${\bf
     M}_{\nu_R}$ to be real, the complex symmetric neutrino mass
   matrix ${\bf M}_{\nu_L}$ has only three independent phase factors
   that come from the parameters $\mu_{2}$, $\mu_{3}$ and
   $\mu_{4}$. Here, to simplify the analysis we will consider the case
   when $\arg \left\{ \mu^{\nu}_{4} \right\} = \arg \left\{
     \mu^{\nu}_{3} \right\}$ or $2\arg \left\{ \mu^{\nu}_{4} \right\}
   = \arg \left\{ 2 \frac{ \left( \mu^{\nu}_{4} \right)^{2} }{
       \overline{M} } + \frac{ \left( \mu^{\nu}_{3} \right)^{2} }{
       M_{3} } \right\}$. The general case, with three independent
   phase factors, will be considered in detail elsewhere.

   In the case considered here the diagonalization of ${\bf
     M}_{\nu_L}$ may be reduced to the diagonalization of a mass
   matrix with two texture zeroes discussed in
   section~\ref{seccion2.4}. The phase factors may be factored out of
   ${\bf M}_{\nu_L}$ as
   \begin{equation}
    {\bf M}_{\nu_L} = 
    {\bf Q}^{\nu} {\bf \cal U}_{_\frac{\pi}{4}} \left( \mu_{_0} { \bf I}_{ _{3 \times 3} } + 
    { \bf \widehat{M} } \right)  
    {\bf \cal U}_{_\frac{\pi}{4}}^{\dagger} { \bf Q}^{\nu}, 
   \end{equation}   
    where ${ \bf Q}^{\nu} = e^{i \phi_{2} } \textrm{diag} \left\{ 1 ,1, e^{i \delta_{\nu} } \right\} $ with 
    $\delta_{\nu} = \phi_{4} - \phi_{2} = \arg \left\{ \mu^{\nu}_{4} \right\} 
    - \arg \left\{ \mu^{\nu}_{2} \right\} $,
   \begin{equation}
    {\bf \cal U}_{_\frac{\pi}{4}} =
    \left( \begin{array}{ccc}
     \frac{1}{\sqrt{2}} & 0 & \frac{1}{\sqrt{2}} \\
     -\frac{1}{\sqrt{2}} & 0 & \frac{1}{\sqrt{2}} \\
      0 & 1 & 0
    \end{array}\right), \quad
    \mu_{_0} = \frac{2 \left| \mu^{\nu}_{2} \right|^{2} }{ \left| \overline{M} \right| } 
    \left( 1 - \left| \lambda \right|  \right),
    \quad \textrm{and} \quad  { \bf \widehat{M} } = 
    \left( \begin{array}{ccc}
     0 & A & 0 \\
     A & B & C \\
     0 & C & 2d
    \end{array} \right)
   \end{equation}
   with $ A = \sqrt{2} \frac{ \left| \mu^{\nu}_{2} \right| \left|
       \mu^{\nu}_{4} \right| }{ \left| \overline{M} \right| } \left( 1
     - \left| \lambda \right|\right) $, $B = \frac{ 2\left|
       \mu^{\nu}_{4} \right|^{2} }{ \left| \overline{M} \right| } +
   \frac{ \left| \mu^{\nu}_{3} \right|^{2} }{ M_{3} } - \frac{ 2\left|
       \mu^{\nu}_{2} \right|^{2} }{ \left| \overline{M} \right| }
   \left( 1 - \left| \lambda \right|\right)$, $ C = \sqrt{2} \frac{
     \left| \mu^{\nu}_{2} \right| \left| \mu^{\nu}_{4} \right| }{
     \left| \overline{M} \right| } \left( 1 + \left| \lambda
     \right|\right) $ and $d = \frac{ 2 \left| \lambda \right| \left|
       \mu^{\nu}_{2} \right|^{2} }{ \left| \overline{M} \right| } $.
   As mentioned before, the diagonalization of ${ \bf M}_{\nu_L}$ is
   reduced to the diagonalization of the real symmetric matrix ${ \bf
     \widehat{M} }$, which is a matrix with two texture zeroes of
   class~I~\cite{Canales:2011ug}.  Hence the matrix ${\bf M}_{\nu_L}$
   is diagonalized by a unitary matrix
   \begin{equation}
   {\bf U}_{\nu} = {\bf Q}^{\nu} {\bf \cal U}_{_\frac{\pi}{4}}{\bf O}^{^{N[I]}}_{\nu} .
   \end{equation}
   In the literature, these similarity  transformations are also known as weak basis transformations, 
   since they leave invariant the gauge currents~\cite{Branco:2007nn}.

   As in the case of the charged leptons, the matrices ${\bf
     M}_{\nu_L}$ and ${ \bf U}_\nu$ can be reparametrized in terms of
   the neutrino masses. For this we use the information that we
   already have about the diagonalization of a matrix with two texture
   zeroes of
   class~I~\cite{Canales:2011ug,Barranco:2010we,Mondragon:1998gy,Mondragon:1999jt}.
   Then, the mass matrix ${\bf M}_{\nu_L}$ for a normal [inverted]
   hierarchy in the mass spectrum takes the form
   {\small \begin{equation} {\bf M}_{\nu_L}^{^{N[I]}} =
       \left( \begin{array}{ccc}
           \mu_{_{0}} + d & d & \frac{1}{\sqrt{2}} \left( C^{^{N[I]}} + A^{^{N[I]}}  \right) \\ \\
           d &  \mu_{_{0}} + d & \frac{1}{\sqrt{2}} \left( C^{^{N[I]}} - A^{^{N[I]}}  \right) \\ \\
           \frac{1}{\sqrt{2}} \left( C^{^{N[I]}} + A^{^{N[I]}} \right)
           & \frac{1}{\sqrt{2}} \left( C^{^{N[I]}} - A^{^{N[I]}}
           \right) & m_{\nu_{1}} + m_{\nu_{2}} + m_{\nu_{3}} - 2
           \left( \mu_{_{0}} + d \right)
    \end{array} \right)
   \end{equation} }
   with {\small $C^{^{N[I]}} = \sqrt{ \frac{ \left( 2d + \mu_{_{0}} - m_{\nu_{1}}  \right) 
    \left( 2d + \mu_{_{0}} - m_{\nu_{2[3]}}  \right) \left(  m_{\nu_{3[2]}} - \mu_{_{0}} -  2d \right) }{2d} }$}
    and
    {\small $A^{^{N[I]}} = \sqrt{ \frac{ \left( m_{\nu_{2}} -  \mu_{_{0}} \right) 
    \left( m_{\nu_{3[1]}} -  \mu_{_{0}} \right) \left( \mu_{_{0}} - m_{\nu_{1[3]}} \right) }{2d} } $}.
   The values allowed for the parameters $\mu_{_{0}}$ and $ 2d + \mu_{_{0}} $ are in the following ranges:
   {\small $ m_{\nu_{2[1]}} > \mu_{_{0}} >  m_{\nu_{1[3]}}$} and 
   {\small $m_{\nu_{3[2]}} >  2d + \mu_{_{0}}  > m_{\nu_{2[1]}}$}.
   The orthogonal matrix ${\bf O}^{^{N[I]}}_{\nu}$ reparametrized in terms of the neutrino masses is given 
   by 
    { \footnotesize \[ 
    \left( \begin{array}{ccc}
     \sqrt{ \frac{ [-1]\left( m_{\nu_{3}}  - \mu_{_{0}}\right) 
       \left( m_{\nu_{2}}  - \mu_{_{0}} \right) f_{1} }{ {\cal D}_{1}^{^{N[I]}} } }   & 
     \sqrt{ \frac{ \left( m_{\nu_{3[1]}}  - \mu_{_{0}} \right) 
       \left(  \mu_{_{0}} - m_{\nu_{1[3]}} \right) f_{2}^{^{N[I]}} }{ {\cal D}_{2}^{^{N[I]}} } } &
     - \sqrt{ \frac{ [-1] \left( \mu_{_{0}}  - m_{\nu_{1}} \right) 
       \left( m_{\nu_{2}}  - \mu_{_{0}}\right) f_{3}^{^{N[I]}} }{ {\cal D}_{3}^{^{N[I]}} } } \\ 
      \sqrt{ \frac{ [-1 ] 2d \left(  \mu_{_{0}} - m_{\nu_{1}} \right) f_{1} }{ 
       {\cal D}_{1}^{^{N[I]}}  } } &   
      \sqrt{ \frac{  2d \left( m_{\nu_{2}}  -  \mu_{_{0}} \right) f_{2}^{^{N[I]}} }{ 
       {\cal D}_{2}^{^{N[I]}}  } } & 
      \sqrt{ \frac{ [-1 ] 2d \left( m_{\nu_{3}}  -  \mu_{_{0}} \right) f_{3}^{^{N[I]}} }{ 
       {\cal D}_{3}^{^{N[I]}}  } } \\
     - \sqrt{ \frac{ [-1] \left( \mu_{_{0}} - m_{\nu_{1}} \right) f_{2}^{^{N[I]}}
       f_{3}^{^{N[I]}} }{ {\cal D}_{1}^{^{N[I]}} } } & 
      \sqrt{ \frac{ \left( m_{\nu_{2}}  - \mu_{_{0}} \right) f_{1}
       f_{3}^{^{N[I]}} }{ {\cal D}_{2}^{^{N[I]}} } } &   
      - \sqrt{ \frac{ \left( m_{\nu_{3}}  - \mu_{_{0}} \right) f_{1}
       f_{2}^{^{N[I]}} }{ {\cal D}_{3}^{^{N[I]}} } } 
    \end{array} \right),
    \]    }
   where :
   { \small \begin{equation}
    \begin{array}{ll}
     {\cal D}_{1}^{^{N[I]}} = 2d \left( m_{\nu_{2}} - m_{\nu_{1}} \right) 
      \left( m_{\nu_{3[1]}} - m_{\nu_{1[3]}} \right), &
     {\cal D}_{2}^{^{N[I]}} = 2d \left( m_{\nu_{2}} - m_{\nu_{1}} \right) 
      \left( m_{\nu_{3[2]}} - m_{\nu_{2[3]}} \right), \\ \\
     {\cal D}_{3}^{^{N[I]}} = 2d \left( m_{\nu_{3[1]}} - m_{\nu_{1[3]}} \right)
      \left( m_{\nu_{3[2]}} - m_{\nu_{2[3]}} \right), & 
     f_{1} = \left( 2d + \mu_{ _{0} } - m_{\nu_{1}}  \right), \\\\
     f_{2}^{^{N[I]}} = [-1] \left( 2d + \mu_{ _{0} } - m_{\nu_{2}}  \right), &
     f_{3}^{^{N[I]}} = [-1] \left( m_{\nu_{3}} - \mu_{ _{0} } - 2d \right).
    \end{array}
   \end{equation} }
 The superscripts $N$ and $I$ denote the normal and inverted hierarchies respectively.
   
  \begin{flushleft}
    { \bf The neutrino mixing matrix}
  \end{flushleft}
   The neutrino mixing matrix ${ \bf V}_{_{PMNS}}$, is the product ${ \bf U}_{eL}^{\dagger} {\bf U}_{\nu} {\bf K}$, 
   where ${\bf K}$ is the diagonal matrix of the Majorana phase factors, defined by 
   ${\bf K}=diag(1,e^{i\alpha},e^{i\beta})$. Now, we obtain the theoretical expression of the elements for the lepton 
   mixing matrix, ${\bf V}^{^{th}}_{_{PMNS}}$,  which is:
   \begin{equation}
    { \bf V}_{_{ PMNS } }^{ ^{th} } = 
    \left(  \begin{array}{ccc}
     V_{ e 1 }^{ ^{th} } & V_{ e 2 }^{ ^{th} } e^{ i \alpha } &  V_{ e 3 }^{ ^{th} }  e^{ i \beta } \\
     V_{ \mu 1 }^{ ^{th} } & V_{ \mu 2 }^{ ^{th} } e^{ i \alpha } &  V_{ \mu 3 }^{ ^{th} } e^{ i \beta } \\
     V_{ \tau 1 }^{ ^{th} } & V_{ \tau 2 }^{ ^{th} } e^{ i \alpha } & V_{ \tau 3 }^{ ^{th} } e^{ i \beta } 
    \end{array} \right) \,
   \end{equation}
   where
   { \small \begin{equation}\label{elemntPMNS}
    \begin{array}{lll}
     V_{ e 1 }^{ ^{th} } = \frac{ \tilde{ m }_{e}  }{ \tilde{m}_{\mu} } O_{11}^{^{N[I]}}
      - O_{21}^{^{N[I]}} e^{ i \delta_{l} },  &
     V_{ e 2 }^{ ^{th} } =  \frac{ \tilde{ m }_{e}  }{ \tilde{m}_{\mu} } O_{12}^{^{N[I]}}
      - O_{22}^{^{N[I]}}  e^{ i \delta_{l} }, & V_{ \tau 1 }^{ ^{th} } =  O_{31}^{^{N[I]}}, \\\\ 
     V_{ e 3 }^{ ^{th} } =  \frac{ \tilde{ m }_{e}  }{ \tilde{m}_{\mu} } O_{13}^{^{N[I]}}
      - O_{23}^{^{N[I]}} e^{ i \delta_{l} },  &   
     V_{ \mu 1 }^{ ^{th} } =  - O_{11}^{^{N[I]}} -\frac{ \tilde{ m }_{e}  }{ \tilde{m}_{\mu} }
       O_{21}^{^{N[I]}} e^{ i \delta_{l} }, &  V_{ \tau 2 }^{ ^{th} } = O_{32}^{^{N[I]}}, \\ \\
     V_{ \mu 2 }^{ ^{th} } = - O_{12}^{^{N[I]}} -\frac{ \tilde{ m }_{e}  }{ \tilde{m}_{\mu} }
      O_{22}^{^{N[I]}} e^{ i \delta_{l} }, &  
     V_{ \mu 3 }^{ ^{th} } = - O_{13}^{^{N[I]}} -\frac{ \tilde{ m }_{e}  }{ \tilde{m}_{\mu} }
      O_{23}^{^{N[I]}} e^{ i \delta_{l} },  &
     V_{ \tau 3 }^{ ^{th} } = O_{33}^{^{N[I]}} 
    \end{array}    
   \end{equation} }   
   with $\delta_{l} = \delta_{\nu} - \delta_{e}$, the elements $O^{^{N[I]}}$ are given in the previous section. 
  
  \begin{flushleft}
   { \bf The Reactor Mixing Angle }
  \end{flushleft}
   The theoretical expression for the lepton mixing angles as functions of the lepton mass ratios are readily 
   obtained when the theoretical expressions for the modulii of the entries in the $PMNS$ mixing matrix, given in 
   eqs.~(\ref{elemntPMNS}), are substituted for $\left|V_{ij}\right|$ in the right hand side of 
   eqs.~(\ref{angulosMezclas}). In a first, preliminary analysis for the reactor mixing angle 
   $\theta_{13}^{ l }$ and for an inverted neutrino mass hierarchy 
   $( m_{ \nu _{ 2 } } > m_{ \nu _{ 1 } } > m_{ \nu _{ 3} } )$ 
   we obtain:
      \begin{equation}
    \sin^{2}{ \theta_{13}^{ l } } \approx  
     \frac{ \left( \mu_{_{0}} + 2d - m_{\nu_{3}} \right) \left( \mu_{_{0}} -  m_{\nu_{3}} \right) }{ 
     \left( m_{\nu_{1}} - m_{\nu_{3}}  \right) \left( m_{\nu_{2}} - m_{\nu_{3}}  \right) }.
   \end{equation}
   Now, with the following values for the neutrino masses $m_{\nu_{2}} = 0.056$~eV, $m_{\nu_{1}}=0.053$~eV and 
   $m_{\nu_{3}}=0.048$~eV,  and the parameter values $\delta_{l}=\pi/2$, $\mu_{0}=0.049$~eV and 
   $d = 8 \times 10^{-5}$~eV, we get 
   $\sin^{2}{ \theta_{13}^{ l } } \approx 0.029  \longrightarrow \theta_{13}^{ l } \approx 9.8^{\circ}$, in good 
   agreement with experimental data~\cite{An:2012eh,Ahn:2012nd}. 
   A more complete analysis,  from a $\chi^{2}$  fit of the exact theoretical expressions for the 
   modulii of the entries of the lepton mixing matrix of the $| ( { \bf V}_{ _{ PMNS } }^{ ^{th} } )_{ij} |$  to the 
   experimental values (for  example the values given in the analysis by Gonzalez-Garcia~\cite{GonzalezGarcia:2007ib}) 
   will be considered in detail elsewhere.
 
  \section{Flavour Changing Neutral Currents (FCNC) and g-2}

  Models with more than one Higgs $SU(2)$ doublet have tree level
  flavour changing neutral currents. In the Minimal $S_{3}$-invariant
  Extension of the Standard Model considered here, there is one Higgs
  $SU(2)$ doublet per generation coupling to all fermions. The flavour
  changing Yukawa couplings may be written in a flavour labelled,
  symmetry adapted weak basis as \cite{Mondragon:2007af,Mondragon:2007nk,Mondragon:2008gm}
  {\small \begin{equation}\label{fcnf-lept}
    \begin{array}{lcl}
     \hspace{-8pt}{\cal L}^{\rm FCNC}_{Y} =
      \left(\overline{E}_{aL} Y_{a b}^{ES} E_{bR} + \overline{U}_{aL} Y_{a b}^{US} U_{bR}
      + \overline{D}_{aL} Y_{a b}^{DS} D_{bR}\right)H_S^0 
      ~+\left(\overline{E}_{aL} Y_{a b}^{E1} E_{bR}+ \overline{U}_{aL} Y_{a b}^{U1} U_{bR} \right. \\ \\ \left.
      +\overline{D}_{aL} Y_{a b}^{D1} D_{bR}\right)H_1^0+
      \left(\overline{E}_{aL} Y_{a b}^{E2} E_{bR}+ \overline{U}_{aL} Y_{a b}^{U2} U_{bR}
      +\overline{D}_{aL} Y_{a b}^{D2} D_{bR}\right)H_2^0+\mbox{h.c.}
    \end{array}
  \end{equation} }
The Yukawa couplings of immediate physical interest in the computation
of the flavour changing neutral currents are those defined in the mass
basis, according to
$\tilde{Y}_{m}^{EI}=U_{eL}^{\dagger}Y_{w}^{EI}U_{eR}$, where $U_{eL}$
and $U_{eR}$ are the matrices that diagonalize the charged lepton mass
matrix defined in eqs.~(\ref{unu}). We obtain~\cite{Mondragon:2007af}
   \begin{equation}\label{y1m}
    \tilde{Y}_{m}^{E1}\approx \frac{m_{\tau}}{v_{1}}\left(
    \begin{array}{ccc}
     2\tilde{m}_{e}  & -\frac{1}{2}\tilde{m}_{e} & \frac{1}{2} x \\ \\
     -\tilde{m}_{\mu} & \frac{1}{2}\tilde{m}_{\mu} & -\frac{1}{2} \\ \\
     \frac{1}{2} \tilde{m}_{\mu} x^2 & -\frac{1}{2}\tilde{m}_{\mu} & \frac{1}{2}
    \end{array} \right)_{m}
    \quad \textrm{and} \quad 
    \tilde{Y}_m^{E2} \approx \frac{m_{\tau}}{v_{2}}\left(
    \begin{array}{ccc}
     -\tilde{m}_{e} & \frac{1}{2}\tilde{m}_{e} & -\frac{1}{2} x \\ \\
     \tilde{m}_{\mu} & \frac{1}{2}\tilde{m}_{\mu} & \frac{1}{2} \\ \\
     -\frac{1}{2} \tilde{m}_{\mu} x^2 & \frac{1}{2}\tilde{m}_{\mu} & \frac{1}{2}
    \end{array} \right)_{m},
   \end{equation}
   where $\tilde{m}_{\mu}=5.94\times 10^{-2}$, $\tilde{m}_{e}=2.876
   \times 10^{-4}$ and $x=m_{e}/m_{\mu}=4.84 \times 10^{-3}$. All the
   non-diagonal elements are responsible for tree-level FCNC
   processes. If the $S_{2}^{\prime}$ symmetry in the Higgs sector is
   preserved~\cite{Pakvasa:1977in,EmmanuelCosta:2007zz}, $\langle
   H_{1}^{0} \rangle = \langle H_{2}^{0} \rangle= v $.

   The amplitude of the flavour violating process $\mu \to 3e$, is
   proportional to $\tilde{Y}_{\mu
     e}^{E}\tilde{Y}_{ee}^{E}$~\cite{Sher:1991km}. Then, the leptonic
   branching ratio,
   \begin{equation}
     Br(\mu \to 3e)=\frac{\Gamma(\mu \to 3e)}{\Gamma(\mu \to e \nu  \bar{\nu})} 
     \quad \textrm{and} \quad
     \Gamma(\mu \to 3e)\approx \frac{m_{\mu}^5}{3\times 2^{10} \pi^3}\frac{\left(Y^{1,2}_{ \mu e}
     Y^{1,2}_{ e e}\right)^2}{M_{H_{1,2}}^4}, 
   \end{equation} 
   which is the dominant term, and the well known expression for 
   $\Gamma(\mu \to e \nu \bar{\nu})$ ~\cite{Yao:2006px}, give 
   \begin{equation} 
    Br(\mu \to 3e)\approx 
    2(2+\tan^2 \beta)^2 \left(\frac{m_e m_\mu}{m_\tau^2}\right)^2\left(\frac{m_\tau}{M_H}\right)^4, 
   \end{equation} 
   where $M_{H}$ is the neutral Higgs involved in the process, whose
   mass we take as $M_H\approx 120~GeV$, and $\tan \beta=1$. We obtain
   $Br(\mu \to 3e)= 2.53 \times 10^{-16}$, well below the experimental
   upper bound for this process, which is $1 \times
   10^{-12}$~\cite{Bellgardt:1987du}.
 
   Similar computations give the numerical estimates of the branching
   ratios for some others flavour violating processes in the leptonic
   sector. These results, and the corresponding experimental upper
   bounds are shown in Table~\ref{table2}. In all cases considered,
   the theoretical estimations made in the framework of the minimal
   $S_3$-invariant extension of the SM are well below the experimental
   upper
   bounds~\cite{Mondragon:2007af,Mondragon:2007nk,Mondragon:2008gm}.

   \begin{table}
    \begin{center}
     \caption{\label{table2}Leptonic FCNC processes, calculated with $M_{H_{1,2}}\sim 120~GeV$.}
    \begin{tabular}{|l|l|l|l|} \hline
     FCNC processes & Theoretical BR &  Experimental  & References \\
     & & upper bound BR & \\ \hline
     $\tau \to  3\mu$ & $8.43 \times 10^{-14}$ & $ 2 \times 10^{-7}$ & 
     B. Aubert {\it et. al.}~\cite{Aubert:2003pc} \\ \hline
     $\tau \to  \mu e^+ e^-$ & $3.15 \times 10^{-17}$ & $2.7 \times 10^{-7} $ & 
     B. Aubert {\it et. al.} ~\cite{Aubert:2003pc} \\ \hline
     $\tau \to \mu \gamma$ &  $9.24 \times 10^{-15}$ & $ 6.8 \times 10^{-8}$ &
     B. Aubert {\it et. al.}~\cite{Aubert:2005ye} \\ \hline
     $\tau \to e \gamma$ & $5.22\times 10^{-16}$ & $ 1.1 \times 10^{-11}$ & 
     B. Aubert {\it et. al.}~\cite{Aubert:2005wa} \\ \hline 
     $\mu \to  3e$ &  $2.53 \times 10^{-16}$ & $  1 \times 10^{-12}$ &
     U. Bellgardt {\it et al.}~\cite{Bellgardt:1987du}  \\ \hline
     $\mu \to e \gamma$ &  $2.42 \times 10^{-20}$ & $ 1.2 \times 10^{-11}$ & 
     M.~L.~Brooks {\it et al.} ~\cite{Brooks:1999pu} \\\hline
    \end{tabular}
    \end{center}
   \end{table}

  \subsection{Muon anomalous magnetic moment}
   In the minimal $S_3$-invariant extension of the Standard Model we
   are considering here,
   the $Z_2$ symmetry decouples the charged leptons from the Higgs
   boson in the $S_3$ singlet representation.  Therefore, at leading
   order of perturbation theory there are two neutral scalars and two
   neutral pseudoscalars whose exchange will contribute to the
   anomalous magnetic moment of the muon~\cite{Mondragon:2007af}.
   Since the heavier generations have larger flavour-changing
   couplings, the largest contribution comes from the heaviest charged
   leptons coupled to the lightest of the neutral Higgs bosons.

   After a straightforward calculation, with the help of (\ref{y1m}),
   we may write $\delta a_{\mu}^{(H)}$ as
   \begin{equation}
   \begin{array}{l}
    \delta a_\mu^{(H)}=\frac{Y_{\mu \tau} Y_{\tau \mu}}{16 \pi^2}\frac{m_{\mu}m_{\tau}}{M_H^2}
    \left(log\left(\frac{M_{H}^2}{m_{\tau}^2}\right)-\frac{3}{2}\right)
    = \frac{m_{\tau}^2}{(246~GeV)^2}\frac{(2+\tan^2\beta)}{32 \pi^2}
    \frac{m_\mu^2}{M_{H}^2}\left(log\left(\frac{M_{H}^2}{m_{\tau}^2}\right)-\frac{3}{2}\right) .
   \end{array}
   \end{equation} 
   Taking again $M_{H}=120~GeV$ and the upper bound for $\tan
   \beta=14$ gives an estimate of the largest possible contribution of
   the FCNC to the anomaly of the muon's magnetic moment $\delta
   a_{\mu}^{(H)}\approx 1.7 \times 10^{-10}$. This number has to be
   compared with the difference between the experimental value and the
   Standard Model prediction for the anomaly of the muon's magnetic
   moment $ \Delta a_\mu=
   a_\mu^{exp}-a_\mu^{SM}$~\cite{Jegerlehner:2007xe}, whose numerical
   value is $ \Delta a_\mu=(28.7 \pm 9.1 )\times 10^{-10}$, which
   means $ \delta a_{\mu}^{(H)}/\Delta a_{\mu}\approx 0.06$.  Hence,
   the contribution of the flavour changing neutral currents to the
   anomaly of the magnetic moment of the muon is smaller than or of
   the order of $6\%$ of the discrepancy between the experimental
   value and the Standard Model prediction.
   
  \section{Conclusions}

  We have discussed the theory of the neutrino masses and mixings as
  the realization of an $S_{3}$ flavour permutational symmetry in two
  models, the Standard Model with an $S_{3}$ flavour symmetry and the
  minimal $S_{3}$-symmetric extension of the Standard Model, with
  three Higgs doublets.
   
  In the Standard Model the imposition of the non-abelian
  permutational symmetry $S_{3}$ as a broken symmetry of flavour,
  leads to a unified treatment of masses and mixings of quarks and
  leptons in which the left-handed Majorana neutrinos acquire their
  masses via the type-I seesaw mechanism. The explicit sequential
  breaking of the $S_{3}$ flavour group according to the chain $S_{3R}
  \otimes S_{3L} \supset S_{3}^{diag} \supset S_{2}^{diag}$, is a
  sufficient condition to define a generic form for the mass matrices
  of all fermions in the theory. In a symmetry adapted or hierarchical
  basis, this generic form is characterized as a mass matrix with two
  texture zeroes of class I. All mass matrices are, then,
  reparametrized in terms of their eigenvalues
  \cite{Mondragon:1998gy,Barranco:2010we,Canales:2011ug,Mondragon:1999jt}.
  After analytically diagonalizing the mass matrices, explicit
  analytical expressions for all entries in the neutrino mixing matrix
  are obtained as functions of the masses of the charged leptons and
  neutrinos and one CP-violating Dirac phase in very good agreement
  with all available experimental data including the recent
  measurements of the reactor angle $\theta_{13}$ made by the T2K,
  Daya Bay and RENO experiments.
   
  In the minimal $S_{3}$-invariant extension of the Standard Model,
  $S_{3}$ is imposed as a fundamental, exact symmetry in the matter
  sector. This assumption leads to extend the concept of flavour and
  generations to the Higgs sector. The fermion sector of the Standard
  Model is left unaltered.  Hence, going to the irreducible
  representation of $S_{3}$, the model has one $SU(2)_{L}$ Higgs
  doublet in the $S_{3}$-singlet representation plus two $SU(2)$ Higgs
  doublets in the two components of the $S_{3}$-doublet
  representation. In this way, all the matter fields, quarks and
  lepton fields, the right-handed neutrino fields, and the Higgs
  fields, belong to the three dimensional ${\bf 1}_{s} \oplus {\bf 2}$
  representation of the group $S_{3}$. A well defined structure of the
  Yukawa couplings is obtained which permits the calculation of mass
  and mixing matrices as functions of the charged leptons and neutrino
  masses~\cite{Kubo:2005sr,Mondragon:2007af}. 
  The magnitudes of the three mixing angles are determined by the
  interplay of the flavour $S_{3}$ symmetry, the charged lepton and
  neutrino mass hierarchies, and the seesaw mechanism. The solar
  mixing angle is almost insensitive to the value of the masses of the
  charged leptons, but its experimental value allowed us to fix the
  scale and origin of the neutrino mass spectrum. The numerical value
  of the atmospheric mixing angle, $\theta_{23}^{l^{th}}$, depends
  strongly on the masses of the charged leptons and is in very good
  agreement with the experiment. In this model the magnitude of the
  reactor mixing angle, $\theta_{13}^{l^{th}}$, is sensitive to the
  difference of the values of the masses of the first and second 
  right-handed neutrinos.  In the case where two of the neutrino
  masses are degenerate, $\theta_{13}$ is different from zero but very
  small \cite{Mondragon:2007af,Mondragon:2007jx}.  Allowing for the
  masses to be non-degenerate gives a values for $\theta_{13}$ in very
  good agreement with recent experimental data.  Explicit expressions
  for the matrices of the Yukawa couplings of the leptonic sector, 
  parameterized in terms of the leptons masses and the VEV's of the
  neutral Higgs bosons in the $S_{3}$- doublet representation, can be
  obtained in this model. Taking for the neutral Higgses $M_{H_{1,2}}$
  a very conservative value $\left( M_{H_{1,2}} \approx 120~ GeV
  \right)$, it is found that the numerical values of the branching ratios
  of the FCNC's in the leptonic sector are well below the
  corresponding experimental upper bounds by many orders of magnitude.
  The contribution of the flavour changing neutral currents to the
  anomaly of the magnetic moment of the muon is small $(6\%)$ but
  non-negligible \cite{Mondragon:2007af,Mondragon:2007nk,Mondragon:2008gm}.

%
%
\section*{Acknowledgements} 
  This work was partially supported by DGAPA-UNAM under contract PAPIIT-IN113712-3 and by CONACyT-
  M\'exico under contract No. 132059.


\providecommand{\WileyBibTextsc}{}
\let\textsc\WileyBibTextsc
\providecommand{\othercit}{}
\providecommand{\jr}[1]{#1}
\providecommand{\etal}{~et~al.}

\end{document}